\documentclass[aps,prx,amsmath,amssymb,reprint,floatfix]{revtex4-2}
\usepackage{amsmath,amssymb}
\usepackage{physics}
\usepackage{graphicx}
\graphicspath{{figures/}}
\usepackage{bm}
\usepackage{soul}
\usepackage{enumitem}
\usepackage[colorlinks=true,linkcolor=blue,citecolor=blue,urlcolor=blue]{hyperref}

\newcommand{\PT}{\mathcal{PT}}
\newcommand{\Top}{\mathcal{T}}
\newcommand{\Pop}{\mathcal{P}}
\newcommand{\bs}{\bm{\sigma}}

\begin{document}

\title{Chirality-induced spin selectivity as a nonequilibrium effect:\\
a unified test of competing mechanisms}

\author{Swagata Acharya}
\email{swagata.acharya@nlr.gov}
\affiliation{National Laboratory of the Rockies, Golden, Colorado, USA}
\author{Mark van Schilfgaarde}
\affiliation{National Laboratory of the Rockies, Golden, Colorado, USA}

\begin{abstract}
For two decades, the origin of Chirality-Induced Spin Selectivity (CISS), the spin polarization of electrons by
nonmagnetic chiral systems without magnetic fields, has remained unsettled. Here, we establish a single exact rule
defining the fundamental conditions for the effect: in nonmagnetic, time-reversal-invariant conductors, all measurable
CISS signals are time-reversal odd and vanish near equilibrium, proving that structural chirality alone is
insufficient. By evaluating leading theoretical mechanisms across four levels of nonequilibrium transport, from coherent
classical driving to fully non-Markovian quantum baths, we verify this selection rule to machine precision. We show that
coherent chiral vibrations generate substantial collinear polarization (up to ~10\% across 0.5--6 THz) that reverses with
handedness, whereas incoherent vibrations yield under 1\%, and non-Markovian bath memory further suppresses the signal.
The chiral geometry first makes the electronic motion chiral by accumulating orbital angular momentum; spin-orbit
coupling (SOC) then converts that into spin. CISS therefore acts as both a spin polarizer and a spin filter, the latter
an order of magnitude weaker. The polarization grows with molecular length and then saturates, matching trends reported
for DNA and peptides.  Reversing the drive converts decaying spin into a handedness-locked charge-current pulse
(inverse-CISS).  We find that once the system is driven, the polarization magnitude scales with an effective spin-orbit
coupling: making heavy atoms and curved light-atom backbones indistinguishable at equivalent effective SOC.  This
enables us to chart how geometry, driving field, and length separate those routes.
\end{abstract}

\maketitle

\section{Introduction and competing ideas}
Electrons transmitted through a chiral system emerge spin polarized, in the absence of any applied field or magnetic
atom; this is chirality-induced spin selectivity
(CISS).~\cite{Naaman2012:JPCL,Naaman2019:NatRevChem,Evers2022} The polarization can reach tens of percent
and is large even in molecules composed of light atoms, where the spin-orbit coupling (SOC) is only a few
meV.~\cite{Xie2011,Eckvahl2023}  Originally discovered as spin-selective electron transmission through organic chiral
molecules,~\cite{Ray1999:Science} CISS has since appeared across chemistry and materials: spin-polarized
photoelectrons,\cite{Gohler2011} enantiomer separation and electrochemical
reactions,~\cite{Mtangi2017,Ghosh2018,Kumar2017,Nair2023,Chae2024,Wang2025} (anomalous) Hall effect,\cite{Kumar2017}
spin-to-charge conversion via inverse CISS,\cite{Sun2024,Dong2025} conductive-probe atomic force
microscopy,\cite{Lu2019} nonreciprocal magnetoresistance (MR),\cite{Qian2022} photoluminescence (PL),\cite{Long2018}
circularly polarized light-emitting diodes,\cite{Hautzinger2024} and optically active
excitons.~\cite{Inui2020,Sercel2025}  The systems range from insulating molecules to high-conductivity crystals.
Chiral molecules are the most common platform; CISS is also seen in metal-halide
perovskites with embedded organic chiral cations (HOIPs), where the chiral molecule's levels sit far from the
Fermi energy and chirality is instead imprinted on the
lead-halide framework by structural distortion.~\cite{Jana2020,Hautzinger2024}

Reviews have mapped the field,~\cite{Michaeli2017,Naaman2019:NatRevChem,Aiello2022:ACSNano,Evers2022,Bloom2024,Dong2025:AnnRevPChem,vanSchilfgaarde2026:Newton}
but no consensus has emerged on the origin of CISS, or on whether one mechanism can break the spin symmetry across
that diversity of compounds and observables.

We show that a single Hamiltonian and symmetry principle can account for a wide range of CISS observations and
address many of the open questions around it.  We treat the chiral conductor as an open quantum system and compare
levels of nonequilibrium theory on that footing, so that we can see what is essential for a signal to appear.
We use two kinds of model tight-binding Hamiltonians, each coupled to a phonon bath.
The questions we set out to answer are:
\begin{enumerate}[topsep=-2pt,itemsep=0pt,leftmargin=12pt]

\item Is CISS a nonequilibrium effect?  In this work, ``nonequilibrium'' refers to response to a perturbation appearing
  outside the linear-response regime where the response can be determined from the ground state.

\item Why is CISS large even in light molecules?  It is widely accepted that CISS vanishes in the absence of SOC, but
  many works predict much smaller polarization than what is observed, especially in light molecules where SOC is very
  small.

\item Is CISS a spin filter (preferential scattering in transport), or spin polarizer (excitations preferentially generate one
  spin)?  Yan has argued that the newer spin polarizer view is strongly favored by recent optical
  experiments.~\cite{Sun2024:SciAdv,Sun2024,Dong2025,Lee2025}

\item Is the symmetry broken in the orbital degrees of freedom, and later transferred to the spin via the SOC, or is the
      SOC somehow directly enhanced by the enantiomer winding?

\item What role do the leads play?  It has been argued that the chiral system inherits large SOC from
      the metal leads, which acts as the seed for the CISS effect.~\cite{Gersten2013,Alwan2021:JACS}

\item Do phonons play a role in CISS?  Phonons, particularly chiral phonons have been shown to contribute to the orbital
      angular momentum,~\cite{Choi2024} and such phonons have been invoked to explain
      CISS.~\cite{Fransson2023:PhysRevRes,Sato2025:arxiv}

\item What is the role of electronic correlations?
      Do they contribute to the CISS effect?~\cite{Huisman2023:JPCC,Xu2024,Chiesa2024:NanoLetters,herbrych2026:arxiv}

\item How to understand inverse CISS in THz experiments?~\cite{Dong2025,Sun2024}  Since the THz frequency exceeds
      characteristic phonon energies, can phonons be excluded as source for CISS?

\item Does CISS appear in a metal? It has been proposed that the observed spin polarization
      is not CISS but a result of electrical magnetochiral anisotropy.~\cite{Zhao2025:NatComm}

\item How to understand temperature dependence of CISS?~\cite{Kiran2016:AdvMater}

\item What distinguishes CISS from the Edelstein effect?

\end{enumerate}
\vspace{12pt}\noindent
Many models have been advanced to explain one or more of these open questions.  Some popular ones are:
\begin{enumerate}[topsep=-2pt,itemsep=0pt,leftmargin=12pt]
\item[$(a)$] \emph{Geometric enhancement of SOC:} The ``atomic'' SOC matrix element is far too small to account for CISS.  A
  ``geometric enhancement'' of SOC was first argued from continuum models,~\cite{Shitade2020,DiVentra2025} and has
  been buttressed by fully relativistic DFT-based transport calculations~\cite{Behera2025,Li2026} that did not assume
  that the SOC originates predominately from a small region near the nucleus (``atomic'' SOC).  Ref.~\cite{Li2026} showed that chirality
  drives a helical charge current. This current makes its own handedness-dependent magnetic
  field. Ref.~\cite{Behera2025} noted, on the other hand, that DFT was too low a level theory to be a quantitative predictor.

\item[$(b)$] \emph{Enhancement from vibrations:} Electron-phonon coupling has often been proposed to boost the
  polarization.~\cite{Fransson2021:Nanolett,Fransson2020:PRB,Du2020:PRB,Zhang2020:PRB,Fransson2022,Vittmann2023,Gupta2025:arxiv} In
  the strong-coupling (polaron) limit the electron drags a phonon cloud and the band narrows, which raises the effective
  spin-orbit splitting relative to the bandwidth.~\cite{KleinMichaeli} A large CISS effect was also found in the weak-coupling limit;~\cite{Fransson2020:PRB} however, a self-consistent study of the same hamiltonian~\cite{Upadhyay2026} finds that the enhancement disappears once the self-energies are made self-consistent.


\item[$(c)$] \emph{Enhancement from coulomb interactions:} Coulomb interactions have been invoked to enhance CISS in
  different ways, e.g.\ by suppressing kinetic energy making the SOC relatively stronger~\cite{Xu2024} or by modifying
  polarization on the acceptor.~\cite{Chiesa2024:NanoLetters}

\item[$(d)$] \emph{Spinterface:} nearly all MR experiments have heavy metal leads.  The spinterface theory hypothesizes
  that leads (with large SOC) amplify SOC in the chiral
  system.~\cite{Gersten2013,Cinchetti2017,Alwan2021:JACS,Pan2025:ACSApplMat,Sarkar2025:ACSNano}

\item[$(e)$] \emph{Orbital angular momentum (OAM):} in this theory, the molecular winding gives rise to a large orbital
  angular momentum, which leaks into the spin sector through local $\mathbf{L}{\cdot}\mathbf{S}$ coupling.  This perspective was argued in
  Ref.~\cite{Gao2022} where large OAM was observed in chiral nanofibers.  The ``orbital Edelstein effect'' was
  discussed by several authors,~\cite{Gao2022,Trama2022,Johansson} and Ref.~\cite{Gobel2025:arxiv} predicts that OAM
  generated purely by geometric chirality is orders of magnitude larger than any direct spin-Edelstein response.

\end{enumerate}
\vspace{12pt}

Each of these explanations likely has a domain of validity, as do other proposals we do not list in detail,
e.g.\ polarization of the acceptor in optically stimulated electron transfer.~\cite{FayLimmer}
What we ask here is whether one Hamiltonian can serve as a common framework for most of the CISS phenomena, and under
what conditions.  We do not claim to settle every mechanism above; we place them side by side on the same
footing so that we can see what each actually does.  Our tight-binding models have a conserving electronic part with
SOC, and a bath that can be coherent or incoherent, with or without memory.

\emph{Time-reversal symmetry must be broken.}  For a conductor with two leads, elastic, unitary,
time-reversal-symmetric transport is enough for a theorem of Bardarson~\cite{Bardarson,YangVanWees} to forbid any collinear spin
polarization of the transmitted current at every energy.  While this theorem has restrictions,~\cite{Sala2026}
reciprocity is inherent to any weakly perturbed hermitian hamiltonian (which has time reversal symmetry $\Top$ by
construction), as was rigorously shown by Kubo.~\cite{Kubo1957:JPSJ}  A ground state hamiltonian that breaks $\Top$
could be posited, but the physical basis for such a hamiltonian is unknown.  Chiral systems' lack of mirror symmetry
breaks parity $\Pop$ but says nothing about $\Top$, so the CISS effect is not explained by chirality alone. $\Top$ can
be broken by pushing the system out of the linear response regime, e.g. through a bias,~\cite{Rikken2001,YangVanWees}
through leakage into a third or magnetic contact,~\cite{Gersten2013} through inelastic
scattering,~\cite{Varela2012_JPCM} or dephasing from the environment.~\cite{GuoSun,Yang2019:PRB,Liu2021:NM,Xu2024:JCTC}
As we show below, CISS vanishes unless $\Pop$ and $\Top$ are both broken.

\section{Model and Hamiltonian}

\begin{figure*}[t]
\centering
\includegraphics[width=\linewidth]{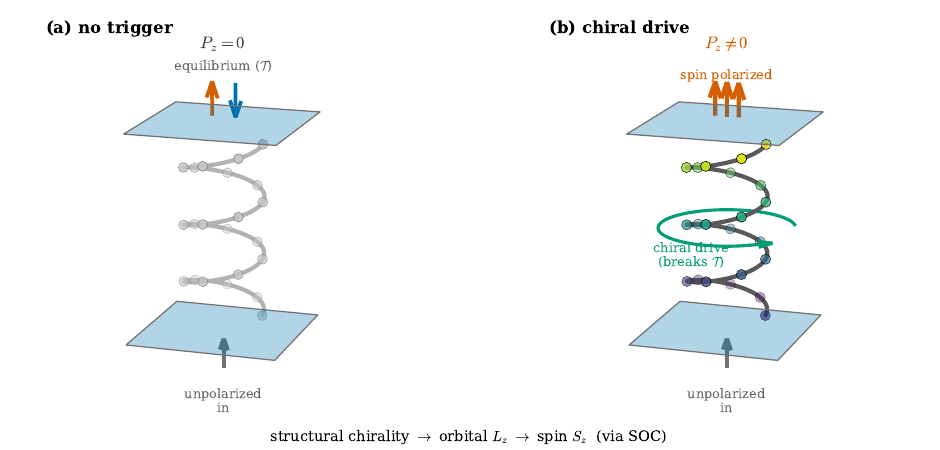}
\caption{Concept. (a) In equilibrium a chiral conductor between two nonmagnetic leads carries no
spin polarization: time-reversal symmetry forces $P_z=0$, and equal up- and down-spin flows cancel.
(b) A chiral drive (a coherent vibration, a bias, or a magnetic contact) breaks time-reversal symmetry
and the transmitted current becomes spin polarized, $P_z\neq0$. Structural chirality produces an
orbital moment $L_z$, which spin-orbit coupling converts into spin $S_z$.}
\label{fig:schematic}
\end{figure*}
We begin with a single-orbital spinful tight-binding Hamiltonian on a helix of $N$ sites
(Fig.~\ref{fig:schematic}), with a fixed geometry (radius $R$, pitch $h$, and $Q$
sites per turn). The electronic part is
\begin{equation}
H_{0}=\sum_{n}\varepsilon_{n}\,c_{n}^{\dagger}c_{n}
-t\sum_{n}\big(c_{n+1}^{\dagger}c_{n}+\text{h.c.}\big),
\label{eq:H0}
\end{equation}
with $c_n=(c_{n\uparrow},c_{n\downarrow})^{T}$, uniform hopping $t$ (our energy unit),
and site energies $\varepsilon_n$ drawn independently and uniformly from $[-W/2,W/2]$
(Anderson disorder of width $W$) whenever we ensemble-average. Chirality enters through a next-nearest-neighbor spin-orbit
hopping of Guo-Sun form,~\cite{GuoSun,GuoSunPNAS}
\begin{equation}
H_{\mathrm{SOC}}=i\lambda\sum_{n}c_{n+2}^{\dagger}\,(\bs\cdot\bm{\nu}_{n})\,c_{n}
+\text{h.c.},\quad
\bm{\nu}_{n}=\frac{\vb{a}_{n}\times\vb{a}_{n+1}}{\abs{\vb{a}_{n}\times\vb{a}_{n+1}}},
\label{eq:HSOC}
\end{equation}
where $\vb{a}_{n}=\vb{r}_{n+1}-\vb{r}_{n}$ is the bond vector and $\bm{\nu}_n$ the chirality-carrying
binormal. The geometry enters the Hamiltonian only through the site coordinates
$\vb{r}_n=(R\cos\phi_n,\,R\sin\phi_n,\,h\phi_n/2\pi)$ with $\phi_n=2\pi n/Q$: these fix the bond
vectors $\vb{a}_n$ and hence the binormals $\bm{\nu}_n$, so $R$, $h$ and $Q$ set the \emph{direction}
of the spin-orbit axis, and its sign fixes the handedness, while $t$ and $\lambda$ remain independent
energy scales. Second-neighbor SOC is needed to distinguish helix handedness.  Reversing handedness flips $\bm{\nu}_n$ and
hence the sign of $H_{\mathrm{SOC}}$. We take SOC corresponding to Pb ($\lambda/t\sim0.3$-$0.6$), which is appropriate
for HOIPs but not for organic molecules. In this minimal single-orbital reduction the only vehicle for carrying angular
momentum in the electronic sector is through $H_{\mathrm{SOC}}$.  This model provides a path to generate spin
polarization, but has no orbital angular momentum.  The $p$-orbital model Eq.~\eqref{eq:orbital} below, removes this
restriction, coupling structural chirality to the \emph{orbital} sector with SOC acting only as the orbital$\to$spin
converter.~\cite{Savi2025,Sala2026} (Orbital-momentum locking survives without SOC.)

For an achiral control we use a straight chain, which has a mirror plane.  Then
$H_{R}=i\alpha\sum_{n}c_{n+1}^{\dagger}\sigma_{y}c_{n}+\text{h.c.}$, and corresponds to a Rashba hamiltonian.
It differs from the chiral term Eq.~\eqref{eq:HSOC} in three ways: it is nearest-neighbor rather than
second-neighbor, its spin axis is fixed ($\sigma_y$) rather than the site-dependent binormal $\bm\nu_n$,
and it survives the mirror plane of the straight chain, so it breaks neither parity nor the handedness
degeneracy. It therefore yields the Edelstein
effect,~\cite{Edelstein,Johansson} a spin polarization \emph{transverse} to the current that is already
present in linear response, in contrast to the collinear, handedness-locked, out-of-equilibrium response
of the chiral term.

\noindent\emph{Phonons.} We treat the chiral phonon at two levels. By a \emph{coherent} phonon we mean a
single-$q$ traveling wave of definite helicity; by an \emph{incoherent} (thermal) phonon we mean the
equilibrium bath in which $+q$ and $-q$ are equally populated. As a \emph{coherent}
classical mode it modulates the bonds as a traveling wave,
\begin{equation}
t_{n,n+1}(\tau)=t+g\cos(\Omega\tau-qn),
\label{eq:drive}
\end{equation}
with classical drive amplitude $g$ (the drive strength; not to be confused with an
electron-phonon matrix element; here the quantized coupling is the matrix $M$, scalar
strength $\chi$, defined below). The phase $qn$ makes the phonon chiral (it carries pseudo-angular momentum;
$q\to-q$ reverses its helicity). We distinguish two senses of ``chiral phonon.''
An \emph{equilibrium} normal mode can already carry angular momentum
$\langle l_z\rangle\neq0$ set by the lattice (indeed the chiral crystal enforces
handedness-locked $l_z$), but such eigenmodes occur in time-reversed pairs
($+l_z$ at $+q$, $-l_z$ at $-q$) that are equally populated in thermal
equilibrium, so the ensemble carries no net angular momentum and preserves
time reversal. What enters Eq.~\eqref{eq:drive} is instead the \emph{coherent},
single-$q$ traveling wave: its helicity is imposed by the excitation (one may
drive $+q$ or $-q$ on the same crystal), so it is independent of the structural
chirality, and being a definite-phase circulating field it is odd under time
reversal ($\Omega\to-\Omega,\,q\to-q$) on its own. Being a scalar (spin-diagonal)
bond modulation, however, it yields a spin response
only in combination with SOC. As a quantized bath we add
\begin{equation}
H_{e\text{-}ph}=\sum_{ij}M_{ij}\,(b+b^{\dagger})\,c_{i}^{\dagger}c_{j}
+\omega_{0}b^{\dagger}b,
\label{eq:eph}
\end{equation}
with a single mode $\omega_0$ and a coupling \emph{matrix} $M_{ij}$ (in site-spin
space) equal to the linear change of the electronic Hamiltonian under the mode
displacement $u$, $M=\chi\,\partial(H_0+H_{\mathrm{SOC}})/\partial u$. It has three
blocks built from the same geometry: a nearest-neighbour hopping modulation
($t_1$, spin-diagonal $\propto\sigma_0$), an on-site Holstein term
($\varepsilon_1$, diagonal $\propto\sigma_0$), and a SOC-modulating piece
($\lambda_1$, next-nearest, $\propto\bm{\sigma}\cdot\bm{\nu}_n$), the last of which is
the only block of $M$ itself that carries a spin-flip operator. The single dimensionless number $\chi$ is the overall
electron-phonon coupling strength, scaling the blocks to the electronic energies
($t_1=\chi t$, $\lambda_1=\chi\lambda$, $\varepsilon_1=\chi\varepsilon_0$); it is the
parameter swept in the self-consistent Born approximation (SCBA) and generalized
Kadanoff-Baym ansatz (GKBA) results below. This does not mean $t_1$ and
$\varepsilon_1$ are inert for spin observables: because $H_0+H_{\mathrm{SOC}}$
already carries static spin--orbit coupling, a purely spin-diagonal modulation
rides on that pre-existing splitting and can drive a spin-dependent signal on
its own, exactly the channel behind the coherent-drive results below, whose
phonon coupling (a scalar bond modulation, Sec.~III) has no $\lambda_1$ term at
all. $\lambda_1$ is an additional, independently present coupling in real
materials, not the sole spin channel; the two can interfere constructively or
destructively depending on parameters. Only if SOC is absent everywhere, in
$H_{\mathrm{SOC}}$ and in $M$, is the phonon truly spin-inert. The coupling $M$ is identical for $+q$ and $-q$: it does not by itself
prefer one helicity, and a thermal bath populates both equally with the Bose factor
$n_B(\omega_0)$, so the incoherent channel carries no net angular momentum. A single helicity
is therefore not selected by $M$; it must be imposed externally or generated dynamically (see Discussion below).

\noindent\emph{Leads.} Two wide-band-limit (WBL) leads couple to the two end sites with rate
$\Gamma$. For the intrinsic (``polarizer'') observable both leads are nonmagnetic and
inject unpolarized carriers. For the magnetoresistance (``filter'') observable one
lead is a spin analyzer, $\Gamma_{L}^{\uparrow,\downarrow}=\Gamma_{0}(1\pm p)$,
and we reverse $p\to-p$ (the experimental magnetization-reversal protocol).

\noindent\emph{Two observables.} (i) The \emph{intrinsic spin polarization} of the
transmitted current, $P_z$, with nonmagnetic leads, from the $2\times2$ outgoing spin density matrix. (ii) The
\emph{magnetoresistance polarization} $P_{\mathrm{MR}}=[I(+p)-I(-p)]/[I(+p)+I(-p)]$. The distinction (spin polarizer
vs.\ spin filter) is itself view of CISS acting as a spin polarizer~\cite{Bloom2024} has been questioned on
  the grounds that it would produce a spin current in equilibrium, and the spin polarizer hypothesis has been
  proposed.~\cite{Wolf2022} Here we report both.

\noindent\emph{A realistic electronic-chirality model}. The minimal model above encodes chirality solely through the SOC
term.  Chirality may also generate a persistent orbital angular momentum (which can be quite large, as noted above), and
subsequently transfer to the spin sector through SOC.~\cite{Savi2025}  We consider a hamiltonian with three $p$ orbitals
per site, two-center Slater-Koster hopping and \emph{atomic} SOC,
\begin{align}
T_{ab}(\vb{d}) &= V_{pp\pi}\,\delta_{ab}+(V_{pp\sigma}-V_{pp\pi})\,d_{a}d_{b},\\
H_{\mathrm{SOC}}^{\mathrm{at}} &= \xi\sum_{n}\mathbf{L}_{n}\cdot\mathbf{S}_{n},
\label{eq:orbital}
\end{align}
where $\vb{d}$ are the bond direction cosines and $\mathbf{L}$ the $l{=}1$ orbital
angular momentum. This is the standard Hamiltonian for real chiral crystals
(tellurium, chiral semimetals),~\cite{Slawinska2023,Roy2022,Yan2024} and it lets us separate \emph{orbital} from
\emph{spin} chirality: with $\xi{=}0$ the helix already produces
orbital-momentum locking $\expval{L_z}(k)$ purely from the winding of the bond
directions, and $\xi$ then converts it to spin-momentum locking
$\expval{S_z}(k)$. In transport we measure both the spin ($P_z$) and the
orbital ($P_{L_z}$) polarization of the transmitted current.

\noindent The two models play distinct roles, and we
keep them separate throughout. The minimal single-orbital model, Eqs.~\eqref{eq:H0}
and~\eqref{eq:HSOC}, carries chirality only through the SOC and has no orbital angular
momentum by construction; we use it for the transport hierarchy and for sorting the
mechanisms, and it is sufficient for every result below that does not require $L$ and
$S$ as independent observables. The $p$-orbital model,
Eq.~\eqref{eq:orbital}, is brought in where the orbital sector is essential:
the equilibrium selection rule (Table~\ref{tab:tparity}) and the orbital-momentum-locking
results (Figs.~\ref{fig:orbital} and~\ref{fig:orbital2}). Table~\ref{tab:summary}
lists which model was used for each demonstration; that column is a bookkeeping of
where the calculation was done, not a claim that the physics is exclusive to one
model. The nonequilibrium selection rule, the coherent-drive hierarchy, and the
handedness and achiral nulls are shared; the $p$-orbital model simply makes the
orbital intermediate visible.

\noindent\emph{Units, and why the numbers are physical.}  Every energy is expressed in units of the hopping $t$, so
$\lambda=0.2$ denotes $\lambda/t=0.2$, and likewise for the SOC, the vibration frequency, the vibration strength, the
lead coupling, and the disorder. Only these ratios matter. Because the hopping itself differs greatly between materials,
the same ratio can be physical in several classes of system. Values chosen below are intended to correspond to realistic
chiral systems.

The spin-orbit ratio $\lambda/t$ spans a wide range across real materials. In lead-halide perovskites the lead $6p$ SOC is large: first-principles
work finds that it splits the conduction band by about $1$~eV.~\cite{Even2013} The bands are also wide, with a hopping
of order $0.5$-$1$~eV, giving $\lambda/t\approx0.1$-$1$. Light molecules occupy the opposite corner yet arrive at the
same place: their atomic spin-orbit coupling is only a few meV,~\cite{Xie2011,Eckvahl2023,KleinMichaeli} but their bands
are narrow. In DNA the base-to-base hopping is only tens of meV,~\cite{EndresDNA} and a few meV divided by tens of meV
again yields $\lambda/t\approx0.1$. A narrow band thus lifts a light molecule into the same ratio as a heavy crystal,
from the opposite direction. Only a wide-band conjugated molecule, with $t$ near an eV, has a truly small
$\lambda/t\approx10^{-3}$. Our runs at $\lambda/t\approx0.1$-$0.6$ therefore describe both perovskites and narrow-band
molecules directly; systems with small $\lambda/t$ require some enhancement of $\lambda$, as we discuss below.

Parameters for vibrations work the same way. Light atoms have stiff bonds and large
zero-point motion. A carbon-carbon stretch is about $0.15$-$0.2$~eV and a
carbon-hydrogen stretch about $0.37$~eV, far above the few-meV octahedral modes of a
soft lead-halide lattice.~\cite{WrightPhonon} In a narrow-band molecule where $t$ is
tens of meV, such a stiff vibration gives $\Omega/t$ of order one, and our value
$\Omega/t\approx0.3$ sits well inside that range. We nonetheless sweep $\Omega$
(Fig.~\ref{fig:freq}) rather than rely on a single point. The vibration amplitude is
physical for the same reason: light atoms undergo large-amplitude motion, and the
hopping decays rapidly with distance, so a single vibration can change it by tens of
percent. This is precisely the polaron regime that molecular charge transport is
known to reach, in which the moving charge drags a cloud of
phonons.~\cite{KleinMichaeli,FayLimmer} The strongest coupling we use, $g/t\approx0.4$, is strong,
but normal for a light-atom molecule rather than an outlier. Weaker couplings are
shown throughout, and the coupling is swept in the magnetoresistance
(Fig.~\ref{fig:summary}), robustness (Fig.~\ref{fig:robust}), and GKBA
(Fig.~\ref{fig:gkba}) results. For a wide-band perovskite all these ratios
shift to smaller values, but the physics is unchanged once the ratios are fixed.

One number is genuinely optimistic. To reach the very smallest $\lambda/t$ of a wide-band molecule, the curvature must
raise the effective coupling by more than tenfold. Such an enhancement rise is consistent with fully relativistic
transport estimates noted above.~\cite{Behera2025,Li2026} Other mechanisms can also enhance the effective SOC, as noted
above, but for narrow-band molecules and the perovskites, no such large enhancement is needed.


\section{A graded sequence of transport calculations}
Table~\ref{tab:levels} summarizes the four treatments we apply. Each isolates a particular way
of breaking time-reversal symmetry, so that a nonzero polarization has an unambiguous
cause.

\begin{table}[t]
\caption{The four transport treatments (Levels~1--4), all in the non-equilibrium
Green's function (NEGF) framework, applied to the \emph{same} Hamiltonian.
``Escape'' names the ingredient beyond the elastic two-terminal no-go.}
\label{tab:levels}
\begin{ruledtabular}
\begin{tabular}{cp{0.38\columnwidth}p{0.32\columnwidth}}
Level & Method & Escape route \\
\hline
1 & Elastic NEGF: elastic, coherent $G^{R}=[E-H-\Sigma_{\rm lead}]^{-1}$ & none (baseline) \\
2 & Floquet-NEGF: coherent classical drive Eq.~\eqref{eq:drive}, exact in $g$, harmonic replicas
 & nonequilibrium: $\Top$ broken by drive\\
3 & SCBA: quantized phonon, self-consistent Born (2nd order in $M$, dressed $G$), Markovian steady state & inelastic, weak coupling\\
4 & GKBA: real-time, non-Markovian collision integral, memory $\tau_c$; Markovian leads & inelastic + bath memory \\
\end{tabular}
\end{ruledtabular}
\end{table}

\noindent\emph{Level 1: Elastic NEGF (baseline).}
$G^{R}(E)=[E-H-\Sigma_{L}^{R}-\Sigma_{R}^{R}]^{-1}$ with leads in the WBL,
$\Sigma_{\alpha}^{R}=-\tfrac{i}{2}\Gamma_\alpha$; transmission
$T=\Tr[\Gamma_{L}G^{R}\Gamma_{R}G^{A}]$. This is elastic and $\Top$-symmetric: by
Bardarson it \emph{must} give $P_z=0$. We use it to fix the units and as the null
against which everything else is measured.

\noindent\emph{Level 2: Floquet-NEGF (coherent chiral phonon).}
The drive Eq.~\eqref{eq:drive} is periodic, so we work in Floquet (harmonic) space,
$[H_F]_{kk'}=H_{k-k'}+k\Omega\,\delta_{kk'}$, with WBL leads on each replica, and
compute the sideband-resolved outgoing spin density matrix. This is \emph{exact in
the drive amplitude} (nonperturbative) and treats the phonon as a coherent field.
The traveling phase breaks $\Top$ dynamically; this realizes the chiral phonon as a
coherent, angular-momentum-carrying field.~\cite{KimSeebeck}

\noindent\emph{Level 3: Self-consistent Born (weak thermal phonon).}
The quantized mode Eq.~\eqref{eq:eph} enters through the Fan-Migdal self-energy at
Born level but computed self-consistently with the dressed Green functions,
\begin{align}
\Sigma^{<}_{\rm ph}(E)&=M\big[n_{B}\,G^{<}(E-\omega_0)+(n_B{+}1)G^{<}(E+\omega_0)\big]M,\\
\Gamma_{\rm ph}&=i(\Sigma^{>}_{\rm ph}-\Sigma^{<}_{\rm ph})\ge0,\quad
\Sigma^{R}_{\rm ph}=\Lambda-\tfrac{i}{2}\Gamma_{\rm ph},
\end{align}
iterated to convergence; the current follows from the Meir-Wingreen generalization of the Landauer
formula.~\cite{MeirWingreen} Here $\Lambda=\mathrm{Re}\,\Sigma^{R}_{\rm ph}$ is the
reactive (principal-value / Hilbert-transform) part of the phonon self-energy,
fixed by the Kramers--Kronig relation to $\Gamma_{\rm ph}$; it shifts levels but
does not itself break the $\Top$ null.  This is the model of
Ref.~\onlinecite{Upadhyay2026}: a \emph{weak, thermal, Markovian-steady} phonon,
and it tests whether coupling to incoherent vibrations can account for CISS.

\noindent\emph{Level 4: Non-Markovian GKBA (bath memory).}
We propagate the single-particle density matrix in real time,
$\dot{\rho}=-i[H,\rho]+I_{\rm lead}(\rho)+I_{\rm ph}(\tau)$, with Markovian
(Lindblad) leads and a two-time phonon collision integral reconstructed via
the GKBA, using a damped phonon correlation $D^{\gtrless}(s)\propto e^{-s/\tau_c}$ of
finite memory time $\tau_c$ ($\tau_c\to0$ Markovian; large $\tau_c$ long memory). It
tests whether \emph{slow} phonon decay lets spin polarization accumulate, the one route
that is invisible to the weak-coupling steady state.

Every level satisfies the expected physics invariants (current conservation
$I_L+I_R=0$, the Keldysh identity $G^{>}-G^{<}=G^{R}-G^{A}$, Floquet unitarity, the
static Bardarson null, the equilibrium $\Top$ null, handedness sign reversal, and the
achiral null) to machine precision (Methods~\ref{sec:methods}).

\section{Results}

\begin{table*}[t]
\caption{Main results by model and transport treatment (Table~\ref{tab:levels}).
The Model column records where the calculation was performed: single-orbital results are not
exclusive to that Hamiltonian; $p$-orbital entries mark cases that require $L$ and $S$ as
separate observables.}
\label{tab:summary}
\begin{ruledtabular}
\begin{tabular}{p{0.48\textwidth}p{0.12\textwidth}p{0.18\textwidth}l}
Result & Model & Treatment & Fig./Tab. \\
\hline
Every measurable CISS observable vanishes at equilibrium ($\Top$-odd selection rule) & $p$-orbital & symmetry + NEGF & Table~\ref{tab:tparity} \\
A coherent chiral vibration gives a handedness-locked $P_z$ that rises with drive frequency, reaching $\sim$10\% near the top of the physical range, and vanishes for a straight chain & single-orbital & Floquet (2) & Figs.~\ref{fig:summary},~\ref{fig:robust} \\
An incoherent vibration gives below a percent, and bath memory only suppresses it further & single-orbital & SCBA (3), GKBA (4) & Figs.~\ref{fig:summary},~\ref{fig:gkba},~\ref{fig:memregime} \\
A signal that rises with temperature is not, by itself, a phonon fingerprint & single-orbital & Floquet (2) & Fig.~\ref{fig:temperature} \\
Chirality produces orbital-momentum locking before SOC; SOC then converts it to spin & $p$-orbital & band structure; Floquet (2) & Figs.~\ref{fig:orbital},~\ref{fig:orbital2} \\
Leakage into a substrate is a distinct route: it polarizes but drains the current & single-orbital & elastic NEGF (1) $+$ loss & Fig.~\ref{fig:leakage} \\
The same mechanism operates in a metal (sizable at all lengths probed) and an insulator (spin-selective tunneling) & single-orbital & Floquet (2) & Fig.~\ref{fig:metalins} \\
The signal is set by the effective SOC alone; enhancement from different sources collapse onto a universal curve & single-orbital & Floquet (2) & Figs.~\ref{fig:routeB},~\ref{fig:freq} \\
Inverse CISS: a decaying spin population drives a handedness-locked charge-current pulse & single-orbital & GKBA (4) & Fig.~\ref{fig:inverse} \\
\end{tabular}
\end{ruledtabular}
\end{table*}

\noindent\emph{Which effects can occur at equilibrium.}
Symmetry alone already fixes much of the picture. In a nonmagnetic chiral
conductor whose Hamiltonian respects time-reversal symmetry, the equilibrium
state does likewise. For any quantity with a definite time-reversal sign,
$\Top\,O\,\Top^{-1}=s_O\,O$, an odd sign forces its equilibrium average to vanish,
\begin{equation}
\expval{O}_{\mathrm{eq}}=\Tr[\rho_{\mathrm{eq}}O]=0 \quad\text{whenever}\quad s_O=-1 .
\end{equation}
We verify the signs on the $p$-orbital model and confirm the zeros to machine
precision (Table~\ref{tab:tparity}). Three consequences follow, none of which
depends on any parameter.
\begin{itemize}[nosep,leftmargin=*]
\item The measured CISS signal is a spin accumulation and is time-reversal odd, so
it vanishes identically at equilibrium. Observing it requires breaking time-reversal
symmetry, through a bias, a drive, or a magnetic contact.
\item Two related quantities do survive at equilibrium: a steady spin current and a
steady orbital current. Both reverse with handedness and vanish for a straight
chain, but they are background flows, not a spin accumulation that can be measured.
\item Under an applied bias the spin accumulation becomes nonzero
(Table~\ref{tab:tparity}, last row), so the measurable effect appears only out of
the linear response regime.
\end{itemize}
This affirms that CISS is an out-of-equilibrium effect. The organizing principle is therefore not a
particular Hamiltonian but a requirement: because the measurable signal is time-reversal odd, it can appear
only once time reversal is broken out of equilibrium. Throughout this work the bulk Hamiltonian stays
Hermitian and time-reversal symmetric, and the breaking is supplied externally in one of three ways: a
nonequilibrium drive (a coherent chiral phonon, the cleanest route), openness (leakage into a substrate or a
third contact, i.e.\ a non-Hermitian loss at the boundary), or a magnetic contact. In the driven and open cases
the Hamiltonian itself is never modified: a biased conductor is the textbook example of a system whose
generator remains completely ordinary, while the \emph{state} it occupies is a nonequilibrium steady state,
sustained indefinitely by unequal lead chemical potentials, with the Hamiltonian remaining in the standard,
Hermitian, time-reversal-symmetric class throughout. The symmetry breaking thus
resides in the drive, the boundary, or the state it produces, not in the closed-system Hamiltonian; this is what distinguishes the
present framework from descriptions that build the breaking into a non-Hermitian bulk Hamiltonian
(see Discussion). A null response should apply to any small perturbation to a
time-reversal-symmetric Hamiltonian, independently of the strength of the SOC.  Indeed, this is a useful way to
discriminate CISS from the Edelstein effect, which is present in linear response.  Usually Edelstein generates spin
polarization transverse to the current, as it does in this model, but it need not be so.~\cite{Taguchi2018:PRB}

Being out of equilibrium is necessary, but it is not the whole story: an ordinary achiral conductor is just as
far out of equilibrium under bias as a chiral one, yet develops no analogous collinear spin accumulation.
What distinguishes the two is a second, independent condition: symmetry. A finite bias supplies the
time-reversal-odd drive; whether that drive is allowed to couple to a $\Top$-odd, parity-even observable such
as $\expval{S_z}$ is fixed by the conductor's point-group symmetry, through Neumann's principle: the
corresponding response tensor must be invariant under every symmetry operation of the material, and any
mirror, inversion, or roto-inversion left in an achiral structure forces this particular tensor component to
vanish identically, at every order in the bias, exactly what we find for the achiral Rashba control, which
develops a current-induced polarization only transverse to the current (the ordinary Edelstein channel that
survives a mirror plane). A chiral conductor, having no improper symmetry operation left to enforce that null,
is the maximal symmetry class in which the collinear channel is allowed; it is not that chiral conductors are
somehow ``more'' out of equilibrium than an achiral wire carrying the same current, but that they are among
the few structures in which the nonequilibrium drive has anywhere to go. This is the same logic, applied here
to a parity-even, time-reversal-odd observable, that governs which crystal classes support a gyrotropic or
kinetic magnetoelectric response more generally.~\cite{Jacquod2012:PRB}

\begin{table}[t]
\caption{Time-reversal parity and equilibrium value of each observable on the
$p$-orbital chiral chain ($N=10$), verified numerically. $\Top$-odd quantities
vanish at equilibrium; $\Top$-even chiral currents do not; the measurable spin
accumulation appears only under bias. The equilibrium rows are evaluated on the
closed system (no leads, from the equilibrium density matrix); the final row (under
bias) uses the open, biased NEGF.}
\label{tab:tparity}
\begin{ruledtabular}
\begin{tabular}{lcc}
observable & $\Top$-parity & $\expval{O}$ \\
\hline
spin density $\expval{S_z}$ (CISS accumulation) & odd & $0$ (eq.) \\
orbital density $\expval{L_z}$ & odd & $0$ (eq.) \\
charge current $\expval{v}$ & odd & $0$ (eq.) \\
spin current $\tfrac12\{v,S_z\}$ & even & $\neq0$ (eq.) \\
orbital current $\tfrac12\{v,L_z\}$ & even & $\neq0$ (eq.) \\
spin density $\expval{S_z}$ under bias & -- & $\neq0$ (non-eq.) \\
\end{tabular}
\end{ruledtabular}
\end{table}

\begin{figure*}[t]
\centering
\includegraphics[width=\linewidth]{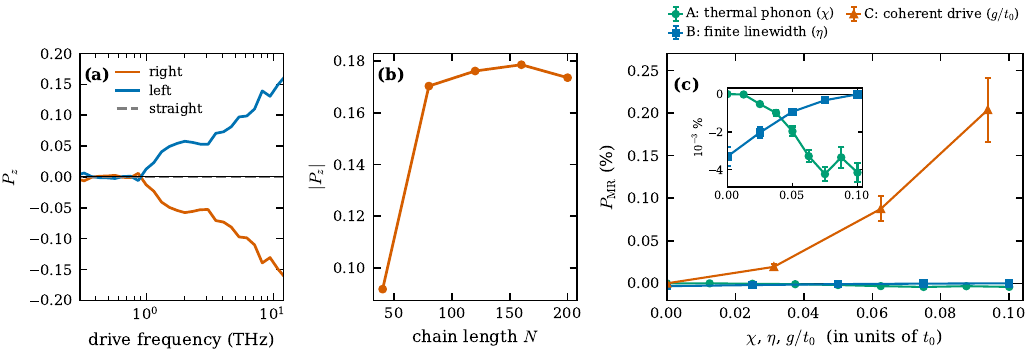}
\caption{Single-orbital model. ($a$) Intrinsic polarization $P_z$ versus drive frequency $\Omega$
(THz scale; the calculation uses $\Omega/t$) for right- and left-handed helices ($N=200$) and a
straight control, from the coherent Floquet drive [Eq.~\eqref{eq:drive}]. ($b$) $P_z$ versus chain
length. ($c$) Magnetoresistance $P_{\mathrm{MR}}=100\,[I(+p)-I(-p)]/[I(+p)+I(-p)]$ (in percent of the mean current at the analyzer lead, polarization $\pm p$) for a thermal phonon versus its coupling $\chi$ (A, SCBA), a thermal phonon with a finite spectral linewidth versus $\eta$ at $\chi=0.1$ (B, SCBA), and the coherent drive versus bond modulation $g/t_0$ (C, Floquet), all in units of the hopping $t_0$ and over the same range.  The inset shows A and B on an expanded scale.  Error bars are statistical, from the disorder average.}
\label{fig:summary}
\end{figure*}

\noindent\emph{A coherent chiral vibration produces the effect.}
Figure~\ref{fig:summary}($a$) shows the polarization $P_z$ against drive frequency for
a right-handed helix, a left-handed helix, and a straight control at the same
SOC. Without the drive the polarization vanishes, as the theorem
requires; with the drive it rises from near zero at the low end of the physical vibration range to about ten
percent near the high end ($0.5$--$6$~THz), and it reverses sign with handedness,
$P_z(\mathrm{R})=-P_z(\mathrm{L})$. The straight control behaves differently: under
the drive it develops a transverse (Edelstein) component but no collinear
polarization, and without the drive it is unpolarized on every axis. A collinear
polarization that appears only for the chiral geometry is thus the clean fingerprint
of the effect.

Panel~($b$) shows that $P_z$ grows with chain length and then plateaus, over the
range $N=40$--$200$ at this SOC and drive strength; Route~B below
(Fig.~\ref{fig:routeB}) shows the same rise-then-saturate shape holds across a
wide range of effective coupling.

\noindent\emph{An incoherent thermal vibration does not.}
Treated instead as a weak thermal bath rather than a coherent drive, the vibration
in the magnetoresistance setup (Fig.~\ref{fig:summary}$c$) yields a spin
polarization of at most $0.005\%$ across the coupling range shown (curve A), and giving
the phonon a finite lifetime does not revive it: curve B is below $0.004\%$ and
falls with the linewidth, reproducing
Ref.~\onlinecite{Upadhyay2026}. The coherent drive (curve C), for the same quantity,
rises with the drive amplitude, to $0.20\%$ at a bond modulation of $9\%$ of the hopping, and so exceeds curve A by more than an order of magnitude.  Larger amplitudes are outside the range of a lattice vibration. In every case the magnetoresistance
(filter) signal is much smaller than the intrinsic (polarizer) signal, consistent
with the filter-versus-polarizer gap observed in
experiment.~\cite{Qian2022,Adhikari2023}

\begin{figure*}[t]
\centering
\includegraphics[width=\linewidth]{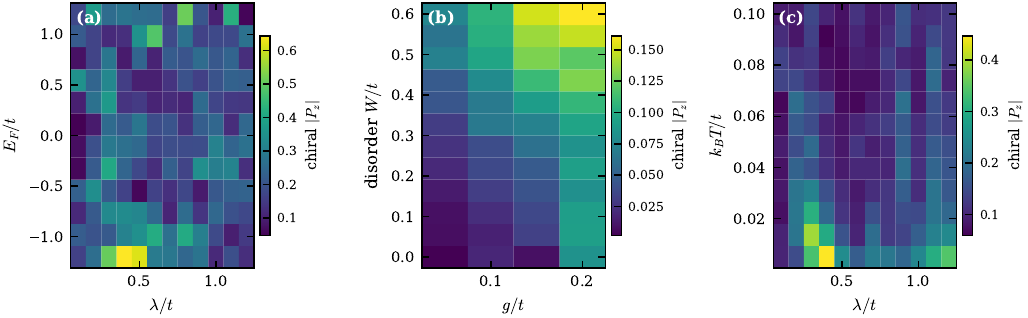}
\caption{Single-orbital model, coherent Floquet drive. Parameter dependence of $P_z$, averaged over
disorder and energy. Left: SOC versus Fermi level. Middle: drive strength versus disorder. Right:
SOC versus temperature at fixed drive. The straight control is unpolarized in every panel.}
\label{fig:robust}
\end{figure*}

\noindent\emph{The effect is robust, not fine-tuned.}
We map the polarization over the relevant parameters, averaged over disorder and
energy (Fig.~\ref{fig:robust}); the straight control remains unpolarized
throughout. Three features stand out: the polarization survives disorder rather than
being suppressed, it grows with the drive strength, and it requires only a finite
SOC. It remains large across the entire scanned range, so no
unusually large coupling is needed once time-reversal symmetry is broken. The
temperature behavior is different, and we treat it next, since that is where
experiments disagree.

\begin{figure*}[t]
\centering
\includegraphics[width=\linewidth]{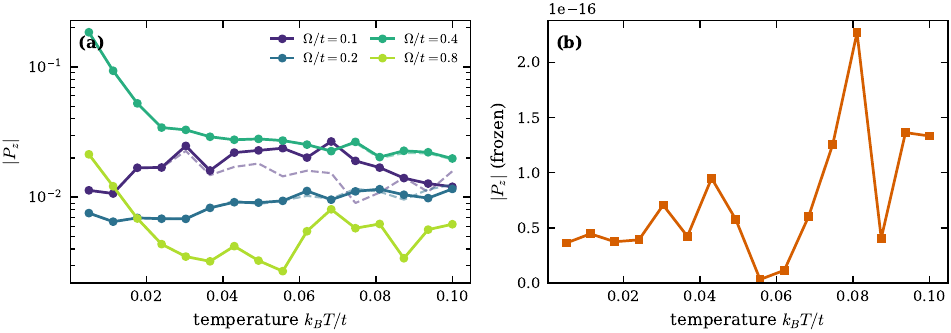}
\caption{Temperature (single-orbital model). Left: coherent Floquet drive, $|P_z(T)|$ for several vibration frequencies, with the
drive scaled by the thermal factor (solid) and held fixed (dashed). Right: a frozen, thermally distorted static lattice (elastic), $|P_z(T)|$.}
\label{fig:temperature}
\end{figure*}

\noindent\emph{Temperature: why experiments disagree.}  The measured temperature trend is not universal. Some
experiments see the signal grow with temperature and read this as a sign of
vibrations,~\cite{Fransson2022,Fransson2023,Kapon2026:arxiv} though a subsequent analysis of the same data finds the
opposite trend.~\cite{Alwan2023} Other work predicts a non-monotonic trend.~\cite{FayLimmer,Vittmann2023} Our model
shows why no single trend is expected. First, a lattice frozen in a thermally distorted geometry still respects
time-reversal symmetry, so its polarization vanishes at every temperature (Fig.~\ref{fig:temperature}, right): the
motion matters, not the distortion \textit{per se}.  Second, once the vibration is dynamic, the temperature trend is set mainly
by the electronic structure rather than by the phonon population; scaling the drive by the thermal factor barely changes
the result (Fig.~\ref{fig:temperature}, left). A signal that increases with temperature is therefore not, by itself,
proof of a vibrational origin. The trend reflects how the thermal window samples the electron transmission, and that
window can rise or fall with the vibration frequency, which is why the reported trends conflict.

\begin{figure}[t]
\centering
\includegraphics[width=\columnwidth]{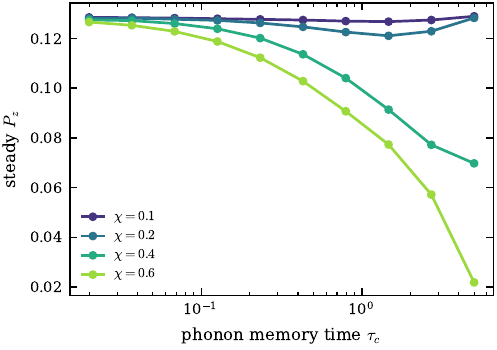}
\caption{Single-orbital model, GKBA, $N=12$. Steady spin accumulation $P_z$ versus vibration memory time
$\tau_c$, at weak and at strong coupling.}
\label{fig:gkba}
\end{figure}

\noindent\emph{Slow-vibration memory only suppresses the effect.}
We now track a different quantity: the spin that accumulates inside the device under
a bias, with no magnet required. This out-of-equilibrium quantity vanishes at
equilibrium and is nonzero under bias, and it is already present with no phonon;
adding a memoryless phonon barely changes it. This quantity and the magnetoresistance
one are thus distinct, and their baselines need not coincide. The relevant question
here is the effect of memory (Fig.~\ref{fig:gkba}): at weak coupling memory has no
effect, while at strong coupling longer memory steadily suppresses the signal. Slow
vibrations therefore do the opposite of accumulating spin.

\begin{figure*}[t]
\centering
\includegraphics[width=\linewidth]{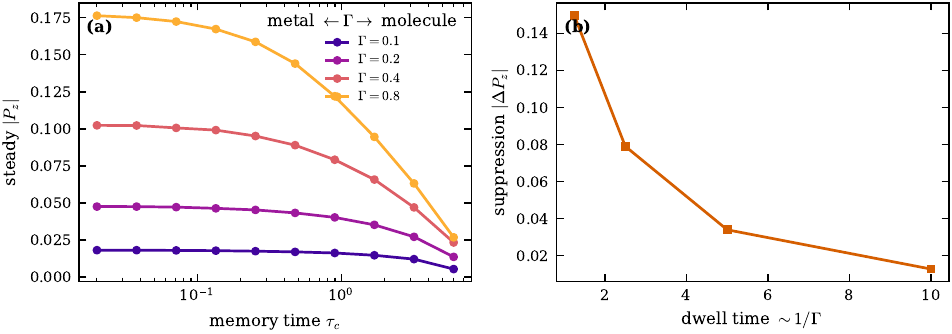}
\caption{Single-orbital model, GKBA, $N=12$, $\chi=0.5$, 32 disorder realizations. Dwell time
$\sim 1/\Gamma$ is the mean residence in the device. Left: steady $|P_z|$ versus memory time $\tau_c$
for large $\Gamma$ (short dwell) to small $\Gamma$ (long dwell). Right: drop in $|P_z|$ from memory,
versus dwell time.}
\label{fig:memregime}
\end{figure*}

\noindent\emph{Does slow-vibration memory help in any regime?}
Memory might be expected to matter more where the electron lingers, as in a
molecule, than in a fast metal. We test this by varying the lead coupling, which
sets the electron dwell time (Fig.~\ref{fig:memregime}). The expectation fails:
memory suppresses the signal in every case, and most strongly where the electron
dwells the shortest time, near where the bare signal is largest. Slow vibrations
hinder the effect everywhere, and the residence-time intuition does not survive the
calculation.

\begin{figure*}[t]
\centering
\includegraphics[width=\linewidth]{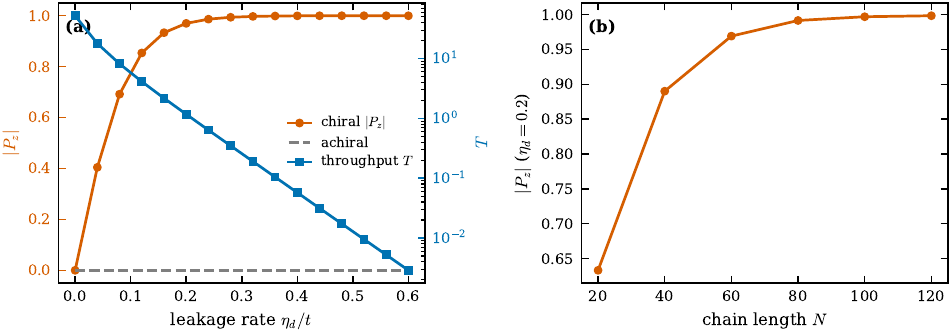}
\caption{Single-orbital model; static elastic NEGF with a uniform on-site loss $\eta_d$, $N=60$, 16 disorder realizations. Left:
chiral $|P_z|$ (red) and the throughput $T$ (blue, log scale) versus leakage rate
$\eta_d$; the straight control is dashed. Right: $|P_z|$ versus length at fixed
leakage.}
\label{fig:leakage}
\end{figure*}

\noindent\emph{Leakage into a substrate is a separate route.}
Not all dissipation suppresses the effect, and two very different processes are
easily confused. The suppression described above applies to scattering that retains
every electron in the system. If instead electrons are allowed to leak out, into a
substrate or a third contact, modeled here by a small loss on every site, the
opposite occurs: the two spins decay at different rates along the chain, so the
surviving current is spin polarized even in static, elastic transport
(Fig.~\ref{fig:leakage}). This is the leakage route of Guo-Sun and
Yang-van~Wees.~\cite{GuoSun,YangVanWees,Evers2022} It vanishes without leakage, requires
both SOC and chirality, reverses with handedness, and approaches
full polarization as the leakage and the length increase. There is a caveat: the
same leak that polarizes the current also drains it, so the current falls rapidly as
the polarization approaches unity.~\cite{Evers2022} Leakage is thus an efficient filter,
but only at a large cost in current. The distinction is clear: ordinary inelastic
scattering suppresses the effect, whereas leakage generates it, and the two must not
be conflated.

\begin{figure*}[t]
\centering
\includegraphics[width=\linewidth]{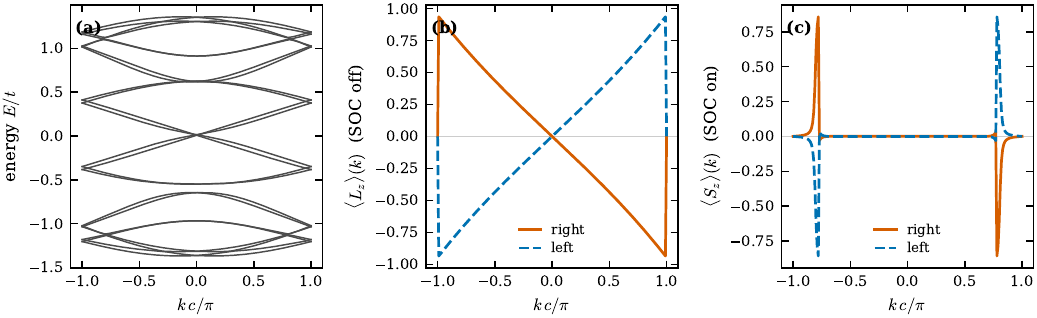}
\caption{$p$-orbital model; equilibrium band structure of the bulk helix (no leads). (a) Bands with SOC on. (b) With SOC off,
the orbital moment $\expval{L_z}(k)$ for right- and left-handed helices. (c) With SOC
on, the spin moment $\expval{S_z}(k)$.}
\label{fig:orbital}
\end{figure*}

\begin{figure*}[t]
\centering
\includegraphics[width=\linewidth]{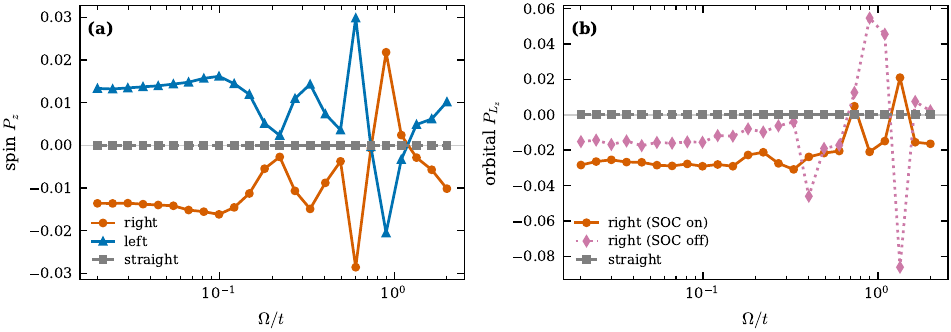}
\caption{$p$-orbital model, transport under the coherent Floquet drive with leads, $N=16$. Left: spin polarization $P_z$ for
right/left helices and the straight control. Right: orbital polarization $P_{L_z}$,
shown with SOC on and off.}
\label{fig:orbital2}
\end{figure*}

\noindent\emph{The chiral geometry polarizes the electrons before SOC acts.}
The $p$-orbital model [Eq.~\eqref{eq:orbital}] shows that the chiral geometry renders
the electrons chiral on its own. With SOC switched off
(Fig.~\ref{fig:orbital}), the filled bands already carry an orbital moment that is
odd in momentum, of order unity, absent for a straight chain, and reversing with
handedness: this is orbital-momentum locking from the geometry alone. Restoring
SOC converts it into the usual spin-momentum locking. The net
equilibrium moments remain zero by time-reversal symmetry, so this too is a scaffold
that requires a trigger. Under the coherent drive (Fig.~\ref{fig:orbital2}) the
current is both spin polarized and orbitally polarized; the orbital polarization
survives even with SOC off, whereas the spin polarization does not.
The current is therefore orbitally polarized simply because the geometry is chiral,
and SOC only converts part of that into
spin.~\cite{LiuYan2021,WanYan2023,Yoda2015} The straight controls remain unpolarized throughout.

\begin{figure*}[t]
\centering
\includegraphics[width=\linewidth]{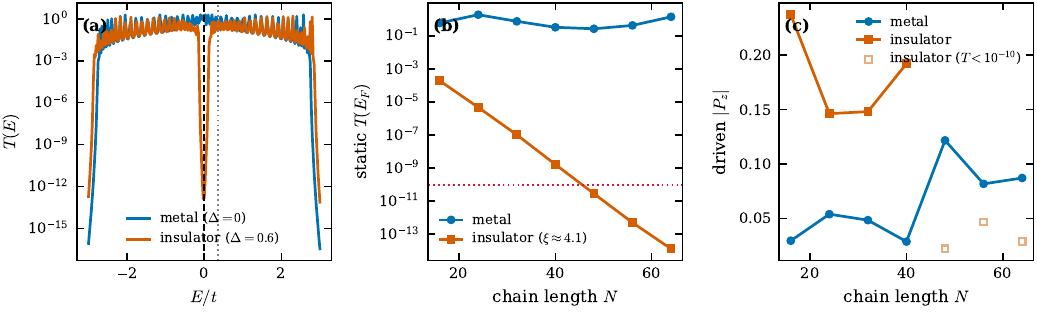}
\caption{Metal versus insulator (single-orbital model); the two regimes differ only by the placement of the
Fermi level, in a band (metal) or in a gap (insulator), not by the lead coupling. Left: transmission versus energy (static elastic NEGF), with the Fermi
level marked in the band and in the gap. Middle: static
transmission versus length. Right: coherent-drive (Floquet) polarization versus length. Open
symbols mark lengths where the current is too small to read.}
\label{fig:metalins}
\end{figure*}

\noindent\emph{Metal versus insulator.}
CISS is measured predominantly in transport, so whether the conductor is a metal or
an insulator is material. In a metal the Fermi level lies within a band; in an
insulator or molecule it lies in a gap. We open a gap in the same chiral chain and
place the Fermi level either in the band or in the gap (Fig.~\ref{fig:metalins}),
from which three points follow. First, the symmetry results are identical in both
cases: without a trigger the polarization vanishes, the straight control remains
unpolarized, and the sign follows the handedness, so the argument does not depend on
metal versus insulator. Second, the magnitude and the length trend differ: in the
metal the transmission is of order unity and nearly length-independent, and the
polarization stays sizable at every length probed, whereas in the insulator the transmission decays
exponentially with length, the signature of tunneling, over a few sites. Third, in
the gap the chiral vibration drives phonon-assisted tunneling: sidebands bridge the
gap, and the surviving current is strongly polarized because the two spins tunnel at
different rates,~\cite{YangVanWees,GuoSun} even though the current itself is tiny.
This is the molecular regime (DNA, peptides),~\cite{Xie2011} whereas the metal case is
the band transport of chiral crystals and perovskites.~\cite{Inui2020,Lu2019} The
same coherent-vibration mechanism operates in both, staying sizable with length in
a metal and turning into spin-selective tunneling in an insulator.  In the metallic case the
electrical magnetochiral anisotropy may be act in parallel with CISS~\cite{Zhao2025:NatComm},
but this result establishes CISS is not extinguished.

\section{Light molecules with weak SOC}
Many molecules that exhibit CISS are composed of light atoms. Carbon, hydrogen,
nitrogen, and oxygen have very weak SOC and should, on their own,
give almost no spin signal, yet experiments on DNA and short peptides observe large
signals. Heavy chiral crystals such as tellurium, and the lead-halide framework of
chiral perovskites, have strong SOC, so a large signal there is
unsurprising; the puzzle is the light molecules.

We test one simple hypothesis: that the relevant SOC is not the
free-atom value but the value set by the curved geometry of the molecule. A tight
helix bends the bonds, mixing the orbitals and raising the coupling, and mechanically
deforming a chiral molecule is known to alter its spin-dependent
transport.~\cite{WangMech,Behera2025} We write the effective coupling as
$\lambda_{\rm eff}=\lambda_{\rm at}(1+\eta_{\rm geo}\kappa)$, where
$\lambda_{\rm at}$ is the weak free-atom value and
$\kappa=|\vb{a}_n\times\vb{a}_{n+1}|/(|\vb{a}_n||\vb{a}_{n+1}|)$ is the local bond curvature,
fixed by the geometry and zero for a straight chain. This multiplicative,
linear-in-curvature form is the one established by curvature-SOC studies of chiral
$sp$ networks;~\cite{Shitade2020,Behera2025,Li2026} it still requires a nonzero atomic
coupling, so it enhances SOC rather than creating it. The magnitude $\eta_{\rm geo}$ is a
phenomenological parameter that we sweep rather than derive; computing it from first
principles for a specific molecule is left to future work.

\begin{figure*}[t]
\centering
\includegraphics[width=\linewidth]{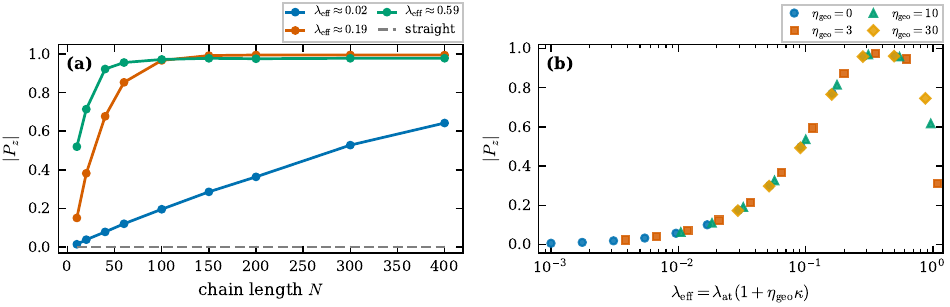}
\caption{Single-orbital model, coherent Floquet drive; curvature-enhanced SOC, averaged over disorder and energy.
(a) Spin polarization versus chain length for a weak, a moderate, and a strong
effective coupling, plus a straight control. (b) The same polarization for many
combinations of free-atom coupling and curvature, plotted against the effective
coupling $\lambda_{\rm eff}$.}
\label{fig:routeB}
\end{figure*}

The result is clean: the signal is controlled solely by $\lambda_{\rm eff}$,
regardless of whether the coupling is large because the atom is heavy or because the
molecule is tightly curved. Figure~\ref{fig:routeB}(b) demonstrates this. Varying
the free-atom value and the curvature over a wide range, all points collapse onto a
single curve when plotted against $\lambda_{\rm eff}$. How much curvature is needed
depends on the band width. For a narrow-band molecule, whose bare ratio
$\lambda_{\rm at}/t$ is already of order $10^{-2}$, a modest enhancement suffices and the
light-atom signal matches the heavy-atom one. For a wide-band molecule, whose bare
ratio can be as small as $10^{-3}$, the same target requires $\eta_{\rm geo}$ of order
tens; this is the genuinely optimistic case, consistent with the order-of-magnitude
enhancements reported in fully relativistic transport studies~\cite{Behera2025,Li2026}
but not something our model predicts on its own.

The signal also grows with the length of the molecule. Figure~\ref{fig:routeB}(a)
shows the spin polarization against chain length: it rises and then saturates, and a
stronger effective coupling saturates sooner, whereas a straight (achiral) chain
gives zero at every length. This reproduces the behavior observed for DNA and
peptides, and even the weak bare coupling reaches a large value in a sufficiently
long molecule.

A vibrating chiral lattice performs the same role. A chiral phonon modulates the
bonds, and this time-dependent modulation of the SOC raises the
effective coupling in the same manner as a fixed curvature; in our calculations it
roughly doubles the signal.

\begin{figure}[t]
\centering
\includegraphics[width=\columnwidth]{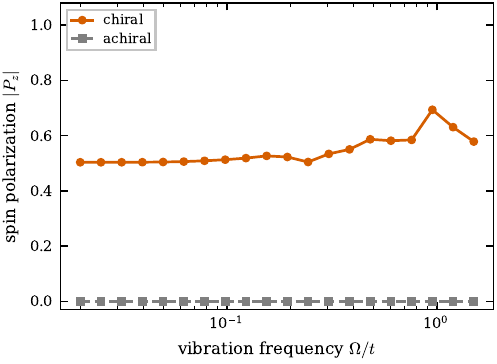}
\caption{Single-orbital model, coherent Floquet drive. Spin polarization versus vibration frequency
$\Omega/t$ for a light molecule with weak coupling ($\lambda_{\rm eff}/t\approx0.1$) and no extra
curvature, averaged over disorder and energy. The straight (achiral) control is zero everywhere.}
\label{fig:freq}
\end{figure}

The effect does not require a special vibration frequency. Sweeping the vibration
frequency for a light molecule with weak coupling and no additional curvature, from
slow to fast bonds and across the entire range of real molecular vibrations, the
spin signal remains large everywhere: it is flat at low frequencies and rises
slightly at the highest ones, while a straight chain remains at zero at every
frequency (Fig.~\ref{fig:freq}). The effect is thus not tied to a single frequency
and holds for the stiff, light bonds of carbon and hydrogen. Real light molecules
also move in many coupled, anharmonic modes simultaneously, which lies beyond our
single clean mode and is left for future work.

\section{Inverse CISS: spin population to charge current}
\begin{figure*}[t]
\centering
\includegraphics[width=\linewidth]{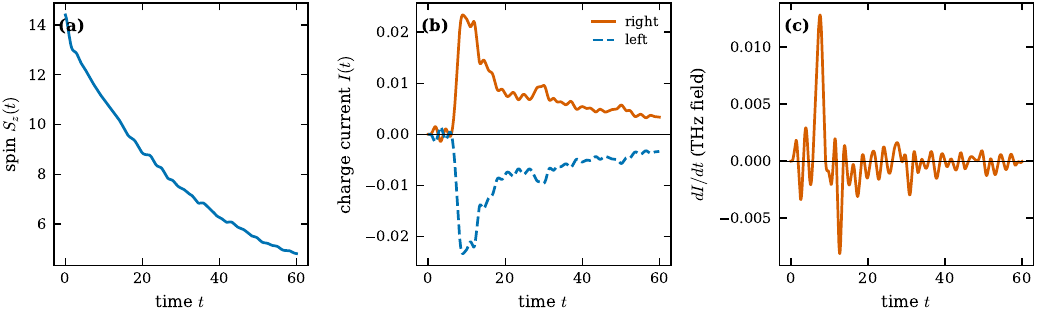}
\caption{Single-orbital model, real-time GKBA, $N=24$, leads at equilibrium. Left:
an injected spin population $S_z(t)$ decaying. Middle: the charge current $I(t)$ it
drives, for right- and left-handed helices. Right: the emitted pulse $dI/dt$.}
\label{fig:inverse}
\end{figure*}

Forward CISS turns a charge current into a spin polarization; the inverse effect,
central to recent ultrafast experiments,~\cite{Dong2025,Sun2024} turns a
non-equilibrium spin population (created e.g.\ by optical orientation) into a charge
current whose time derivative radiates terahertz. We compute it directly in real time
(GKBA) with the leads held at equilibrium half-filling, with \emph{no} charge bias, and an
initial \emph{pure spin population with zero charge imbalance}
(Fig.~\ref{fig:inverse}). As the injected spin relaxes, the chiral$+$SOC structure
rectifies it into a \emph{handedness-locked} through charge current $I(t)$ that peaks on
the spin-relaxation timescale and integrates to a net transferred charge; the emitted
field $\propto dI/dt$ is a bipolar transient. The signal vanishes without SOC and for
the achiral control, flips sign exactly with handedness, and is zero if the spin
population starts at equilibrium, so inverse CISS is the exact reciprocal of forward
CISS and, like it, an intrinsically non-equilibrium, $\Top$-odd response. This provides
a microscopic picture of the inverse-CISS terahertz emission reported in chiral
perovskites~\cite{Dong2025}: no ingredient beyond chirality, SOC, and a relaxing
spin population is required. That the emitted radiation lies in the terahertz band, above characteristic
phonon energies, does not exclude a phonon origin. The radiated field tracks $dI/dt$ of a \emph{rectified}
charge transient whose duration is set by the spin-relaxation time, not by any phonon frequency, so its
spectrum reaches into the terahertz regardless of the underlying mode; and, conversely, low-frequency
chiral phonons can themselves be launched impulsively by an optical or terahertz pump, so a terahertz
probe does not require a terahertz-energy phonon. The relative energy scales are thus not a basis on which
to rule phonons out (the same caveat applies to frequency-based exclusions of phonons in related
magneto-chiral terahertz emission from photoexcited tellurium).\cite{Huang2026:NatPhys}

\section{Discussion: sorting the competing ideas}
We can now place each proposal against the no-go rule.
\begin{itemize}[topsep=-2pt,itemsep=0pt,leftmargin=12pt]
\item \emph{Coherent chiral vibration.} This is the mechanism our calculations support:
it produces a large signal that reverses with handedness, vanishes for a straight
chain, and grows with length.
\item \emph{Weak or incoherent vibration}.~\cite{Upadhyay2026,Fransson2022,Fransson2024}
This produces almost nothing. It vanishes in the memoryless steady state, as found
by others, and slow-vibration memory only weakens it further, with the suppression
largest in the fast (metal-like) case rather than the slow one; the expectation that
long molecules allow memory to help is thus incorrect. The temperature result is
consistent: scaling the drive by the thermal factor barely changes the signal, so a
signal that increases with temperature is not, by itself, evidence of a vibrational
origin.
\item \emph{Inverse effect (spin to charge)}.~\cite{Dong2025,Sun2024} We find
it to be the exact reverse of the forward effect: a decaying spin population drives
a charge-current pulse that reverses with handedness and radiates terahertz through
$dI/dt$. It vanishes without SOC, for a straight chain, and at
equilibrium, and requires nothing beyond chirality, SOC, and an
out-of-equilibrium spin population.
\item \emph{Leakage into a substrate}.~\cite{GuoSun,YangVanWees} This is a separate
route: electrons leak out, the two spins leak at different rates, and the
surviving current is spin polarized. It is thus an independent mechanism for CISS, subject to the caveats noted earlier.
\item \emph{Enhanced effective SOC}.~\cite{Behera2025,Wolf2022} Even large spin-orbit
coupling gives zero in the elastic two-terminal limit; SOC sets the
magnitude of the effect but is not itself the mechanism. Several possibilities for enhancing SOC were noted earlier.
Geometric enhancement is one very likely route;~\cite{Shitade2020,DiVentra2025,Behera2025,Li2026} other contributors can
be coulomb correlations, \cite{Xu2024,herbrych2026:arxiv} and inheriting SOC from heavy metal leads.~\cite{Gersten2013,Alwan2021:JACS}
The model suggests it may be difficult to distinguish the possible enhancement mechanisms on experimental grounds,
as the enhancement collapses onto a universal curve (Fig.~\ref{fig:routeB}).
\item \emph{Self-generated field}.~\cite{Li2026} A steady equilibrium helical
current is time-reversal odd and therefore forbidden in a nonmagnetic ground state.
If instead the current is driven, it is an out-of-equilibrium effect, consistent
with the coherent route. The proposed self-field of about $0.1$~T gives a
Zeeman energy far below $k_BT$, which suggests any signal must come from the current-borne
spin rather than from Zeeman splitting.  That check remains to be done.
\item \emph{Orbital chirality.} We find this to be real and to precede spin-orbit
coupling: the chiral geometry produces orbital-momentum locking and an
orbital-polarized current even with SOC off, and spin-orbit
coupling then converts part of it into spin. This is a genuine electronic route,
distinct from vibrations, but like the others it vanishes at equilibrium and
requires a trigger to become a current. It is a scaffold, not a mechanism on its
own.
\item \emph{Non-Hermitian / $\PT$-symmetric descriptions.} A non-Hermitian exchange ($\PT$-symmetric, but
odd under parity and time reversal separately) has been proposed as an intrinsic, equilibrium origin of
CISS.~\cite{TheilerBeard1,Theiler2026} That construction places time-reversal breaking inside the
closed-system Hamiltonian. We reject that placement: a Hermitian, $\Top$-symmetric generator forbids a
nonzero $\Top$-odd equilibrium signal, and inserting an explicit $\Top$-odd term into an equilibrium
Hamiltonian removes the system from that class, with the attendant violations of microscopic reversibility
and Onsager reciprocity that equilibrium thermodynamics requires. A $\PT$-symmetric non-Hermitian Hamiltonian is not
time-reversal symmetric, so it can support a nonzero $\Top$-odd signal without a drive precisely because the breaking
has been built into the generator.  Our framework keeps the bulk Hamiltonian Hermitian and $\Top$-symmetric and breaks
time reversal only out of equilibrium, through the drive, the open boundary, or the state those produce. Non-Hermitian
Hamiltonians remain useful as effective open-system descriptions; but absent evidence for a non-Hermitian, $\PT$
symmetric hamiltonian in equilibrium, the microscopic origin of any such effective term is, in our account, the
out-of-equilibrium conditions the experiments impose.
\end{itemize}

\medskip\noindent\emph{Magnetoresistance.} Every magnetoresistance measurement compares currents for
opposite lead magnetizations, $P_{\mathrm{MR}}=[I(+p)-I(-p)]/[I(+p)+I(-p)]$, and therefore requires a
magnetic (or otherwise spin-selective) contact in addition to the chiral conductor. In our calculations
$P_{\mathrm{MR}}$ is handedness-locked and reverses with $p$, but is roughly an order of magnitude smaller
than the intrinsic polarization $P_z$ (Fig.~\ref{fig:summary}$c$). The reason is that the filter
signal is second order: it needs both the analyzer magnetization and the out-of-equilibrium drive, whereas
$P_z$ needs the drive alone. This ordering, filter $\ll$ polarizer, is the one seen in
experiment,~\cite{Qian2022,Adhikari2023} and it cautions against reading a small magnetoresistance at zero bias as evidence for a weak intrinsic effect.

\medskip\noindent\emph{Why the phonon must be quantized.} A static lattice, whether at the minimum energy structure or taken as a snapshot of the lattice distorted by thermal fluctuations, preserves
time-reversal symmetry and gives no signal at any temperature (Fig.~\ref{fig:temperature}, right). A
classical mean-field treatment of the vibration, which only shifts the bonds, is of the same character.
The thermal channel produces a signal only once the mode is quantized, because the quanta supply real
absorption and emission at $\pm\omega_0$: these inelastic events exchange energy with the bath, break
detailed balance, and open the $\Top$-breaking channel that an elastic or adiabatic description lacks.
Quantization is thus no mere technical refinement, but the physical origin of the inelastic route; freezing
the displacement or taking $\omega_0\to0$ restores the null. In this sense the classical (mean-field)
treatment of the \emph{thermal} bath is effectively adiabatic and elastic: a static or slowly varying bond
shift only renormalizes the levels and, in the adiabatic limit, exchanges no energy with the electrons, so
it cannot break detailed balance. The coherent Floquet drive is the opposite case: although it too is a
classical field, it is strongly time dependent and breaks time reversal through its circulating phase, which
is why it produces a signal without being quantized. Quantization matters specifically for the
\emph{incoherent} thermal phonon, which has no such coherent phase and can break detailed balance only
through real emission and absorption of quanta.

\medskip\noindent\emph{Selection of a single phonon helicity.} The coherent route assumes one traveling
chiral mode of definite $q$. In thermal equilibrium the $+q$ and $-q$ modes are degenerate and equally
populated, so the crystal alone provides no net helicity, and how a single helicity is selected in a given
experiment is an open question. Two routes are physical. Externally, chiral phonons of definite
pseudo-angular momentum can be launched selectively, for example by circularly polarized or terahertz
pumping,~\cite{Choi2024} as now observed directly in several crystals.~\cite{Zhu2018,Ueda2023,Tauchert2022}
Intrinsically, the electron system can feed back on the bath: a chiral electronic current carries angular
momentum and can pump phonon angular momentum~\cite{Yokoyama2025,Yokoyama2026}, preferentially driving one helicity and selecting it
self-consistently. That intrinsic route, current-induced selection of one helicity
through electron--phonon feedback, is established in Ref.~\cite{Acharya2026Selection}, which derives
the selection from transport rather than assuming a single helicity as an
input.

\medskip\noindent\emph{Scattering and orbital accumulation.} Our clean single-mode model omits the
elastic and inelastic scattering that, in a real molecule, would relax the accumulated orbital moment.
Since a large orbital angular momentum has been measured in chiral systems,~\cite{Gao2022} these scattering
corrections to $\expval{L_z}$ may be quantitatively important; they could reduce the accumulation or, through
orbital-texture scattering, redistribute it. A full account is beyond what we do here. We note only
that the geometric orbital-momentum locking that seeds the effect is a band property and survives weak
disorder (Fig.~\ref{fig:robust}).

\section{Conclusion}
CISS looks like a tangle of competing mechanisms, but one exact rule organizes it for us: in a nonmagnetic conductor
with time-reversal symmetry, the spin polarization vanishes at equilibrium, so every observable signal is
out-of-equilibrium. Each proposal is then simply a route for breaking time-reversal symmetry: a bias, a coherent
drive, leakage, or a term added to the Hamiltonian by hand. We tested those routes on one tight-binding Hamiltonian and
asked under what conditions a sizable polarization appears.

We find that a coherent chiral vibration is the cleanest trigger we tried: it gives a large signal that reverses with
handedness and vanishes for a straight chain. An incoherent vibration does not, and slow vibrations weaken the signal
rather than strengthen it. Bath memory suppresses rather than builds the response, so CISS is not, in our calculations,
a non-Markovian effect: a long-lived vibrational memory dephases the accumulating spin instead of sustaining it. The
chiral geometry makes the electronic motion chiral even with spin-orbit coupling off; SOC then converts that motion into
spin. Leakage into a substrate is a separate route that can produce a signal but drains the current, and it should not
be confused with ordinary inelastic scattering, which works the other way.  Run in reverse, the same model turns a
decaying spin population into a charge-current pulse. We also find both a spin-polarizer piece (intrinsic $P_z$) and a
weaker spin-filter piece (magnetoresistance), so the two experimental descriptions are limiting views of the same response.

The Hamiltonian is the same for light molecules and heavy chiral crystals; only the amplification differs. Tellurium
and the lead-halide framework of chiral perovskites already have strong SOC; light molecules do not, but geometric
enhancement (and possibly Coulomb interactions~\cite{herbrych2026:arxiv}) can raise the effective coupling toward the heavy-atom range. The
signal then grows with molecular length and saturates, as experiments find.

A few conclusions follow from symmetry alone and do not depend on our models: chirality by itself does not produce the
effect; a signal that rises with temperature is not, by itself, evidence of vibrations; incoherent phonons do not give
a significant result; and slow vibrational memory cannot rescue an incoherent phonon.

Our models remain one-dimensional helices: a minimal spinful chain and a three-orbital ($p_x,p_y,p_z$) Slater-Koster
Hamiltonian with atomic SOC. Ref.~\cite{Acharya2026Selection} takes the next step towards a fully first principles description:
it shows that a dc transport current itself selects the phonon helicity in elemental tellurium, with electronic
structure, phonons, and electron--phonon coupling computed from first principles, so the coherent chiral mode is
an output of the transport rather than an assumed input.

\section{Methods: models, self-energies, and invariant checks}
\label{sec:methods}
The four levels are nonequilibrium Green's function (NEGF) treatments of a central region (the
helix) placed between two leads; for the formalism see Refs.~\cite{Datta1995,HaugJauho,Stefanucci2013}.
Unless stated otherwise the central Hamiltonian $H$ is the single-orbital model, Eqs.~\eqref{eq:H0} and
\eqref{eq:HSOC}, with its full SOC; the $p$-orbital model, Eq.~\eqref{eq:orbital}, is used only for
Figs.~\ref{fig:orbital}--\ref{fig:orbital2} and Table~\ref{tab:tparity}. All levels share the same leads,
described in the wide-band limit (WBL) by an \emph{energy-independent} self-energy
$\Sigma^{R}_{\alpha}=-\tfrac{i}{2}\Gamma_{\alpha}$ ($\alpha=L,R$). This is a deliberate simplification: a
realistic lead self-energy is frequency dependent and carries its own band edges and spin-orbit coupling,
which would shift magnitudes and could add a spinterface contribution. We adopt the WBL because our
conclusions rest on symmetry (the equilibrium $\Top$ null, the handedness reversal, the achiral null), which
the WBL respects; it is not intended as a quantitative model of a specific lead. $\Gamma_{\alpha}$ is a fixed
coupling strength, independent of energy, bias, or time; the bias enters separately, only through the chemical
potentials $\mu_{L,R}$ inside the lead Fermi functions that multiply $\Gamma_{\alpha}$ in $\Sigma^{\lessgtr}_{\alpha}$.
A finite $\Gamma_{\alpha}$ is in fact what guarantees a well-defined, time-translation-invariant steady state
exists at all: it pushes the poles of $G^{R}$ off the real axis, so that any transient dependence on how the
leads were connected decays away, leaving the stationary, energy-domain solution used throughout.

\emph{Elastic NEGF (Level 1)}~\cite{Datta1995}.
$G^{R}(E)=[E-H-\Sigma^{R}_{L}-\Sigma^{R}_{R}]^{-1}$, transmission
$T(E)=\Tr[\Gamma_{L}G^{R}\Gamma_{R}G^{A}]$; the outgoing $2\times2$ (single-orbital) or $6\times6$
($p$-orbital) spin(-orbital) density matrix is $\rho_{\rm out}=t\,t^{\dagger}$,
$t=\sqrt{\Gamma_{R}}\,G^{R}\sqrt{\Gamma_{L}}$. The leakage results (Fig.~\ref{fig:leakage}) add a uniform
on-site loss $-\tfrac{i}{2}\eta_d$ at every site.

\emph{Floquet-NEGF (Level 2)}~\cite{Aoki2014}. The coherent chiral phonon enters as the periodic,
\emph{undamped} bond modulation Eq.~\eqref{eq:drive}, whose only time dependence is the drive phase (it does
not decay or scatter). With $H(\tau)=H_{0}+H_{+1}e^{-i\Omega\tau}+H_{-1}e^{+i\Omega\tau}$ the Floquet
Hamiltonian is $[\mathcal H_{F}]_{kk'}=H_{k-k'}+k\Omega\,\delta_{kk'}$ with WBL self-energies on each replica;
$\rho_{\rm out}=\sum_{k}t^{(k)}t^{(k)\dagger}$ summed over sidebands. The treatment is nonperturbative in the
drive amplitude $g$. Sparse and dense solvers agree to machine precision.

\emph{Self-consistent Born (Level 3)}~\cite{Frederiksen2007,Galperin2007}. The vibration is a single quantized
Einstein mode $\omega_0$ coupled to the electrons by Eq.~\eqref{eq:eph} through the matrix
$M=\chi\,\partial(H_0+H_{\rm SOC})/\partial u$, the linear change of $H$ under the mode displacement $u$ (in
the single-orbital model the spin-flipping block is second-neighbour, inherited from the Guo-Sun SOC). The
Fan-Migdal self-energy is taken at Born level but evaluated self-consistently with the dressed $G$,
$\Sigma^{<}_{\rm ph}(E)=M[n_B G^{<}(E-\omega_0)+(n_B{+}1)G^{<}(E+\omega_0)]M$, $\Sigma^{>}_{\rm ph}$ analogous,
$\Gamma_{\rm ph}=i(\Sigma^{>}_{\rm ph}-\Sigma^{<}_{\rm ph})$, $\Sigma^{R}_{\rm ph}=\Lambda-\tfrac{i}{2}\Gamma_{\rm ph}$;
the current follows from the Meir-Wingreen formula~\cite{MeirWingreen}
$I_{L}=(2\pi)^{-1}\!\int\!dE\,\Re\{i\Tr[\Gamma_{L}(G^{<}+f_{L}(G^{R}-G^{A}))]\}$. The ``finite-lifetime''
variant (curve B of Fig.~\ref{fig:summary}$c$) gives the mode a finite spectral linewidth $\eta$ (a broadened
phonon propagator) as a steady-state non-Markovian proxy.  Curves A and B are for a chain of $N=24$ sites averaged over 256 disorder samples; the self-consistent iteration is continued until $P_{\mathrm{MR}}$ no longer changes.

\emph{GKBA (Level 4)}~\cite{Lipavsky1986,Stefanucci2013}. We propagate the single-particle density matrix in
real time, $\dot\rho=-i[H,\rho]+I_{\rm lead}(\rho)+I_{\rm ph}(\tau)$, using the \emph{same} $H$ (including SOC)
and the same leads as above, with the leads in a Markovian (Lindblad/WBL) form. The term $I_{\rm ph}$ is the
\emph{electron}-phonon collision integral generated by the harmonic coupling Eq.~\eqref{eq:eph} (electrons
scattering off the mode, not a phonon-phonon/anharmonic term), reconstructed from the single-time $\rho$ via
the GKBA together with a two-time phonon correlation $D^{\gtrless}(s)\propto e^{-s/\tau_c}$ of finite memory
$\tau_c$ (truncated at $s>14\tau_c$). Here $\tau_c$ is the phonon's own correlation/decay time, physically
set by anharmonicity and coupling to other baths and parameterized directly here, and $\tau_c\to0$ recovers the
Markovian limit.

\emph{Invariant checks (gate every production run).} Current conservation
$I_L+I_R=0$; Keldysh identity $G^{>}-G^{<}=G^{R}-G^{A}$; Floquet spectral-identity
unitarity $i(G_F-G_F^{\dagger})=G_F\Gamma_F G_F^{\dagger}$; static Bardarson null;
sparse$=$dense; $g\!\to\!0$ and $\Omega\!\to\!0$ limits; GKBA energy/trace
conservation and physical $\rho$ (eigenvalues in $[0,1]$); equilibrium $\Top$-odd
observables vanish (Table~\ref{tab:tparity}); handedness sign-flip; achiral null.
All hold to machine precision.

\section*{Data Availability}
All data supporting the findings of this study will be provided by the
corresponding author upon reasonable request.

\section*{Author Contributions}
S.A.\ conceived the study, developed the codes, performed all calculations, and wrote the paper.
M.v.S.\ contributed to the writing and editing of the draft.

\begin{acknowledgments}

This work was authored by the National Laboratory of the Rockies for the U.S.
Department of Energy (DOE) under Contract No.\ DE-AC36-08GO28308.  Funding was provided by the Computational Chemical Sciences program
within the Office of Basic Energy Sciences, U.S.\ Department of Energy.  SA acknowledges the use of the National Energy
Research Scientific Computing Center, under Contract No.\ DE-AC02-05CH11231 using
NERSC award BES-ERCAP0021783, and also acknowledges that a portion of the research
was performed using computational resources sponsored by the Department of
Energy's Office of Energy Efficiency and Renewable Energy and located at the
National Laboratory of the Rockies.  The views expressed in the article do not
necessarily represent the views of the DOE or the U.S. Government.  The U.S.
Government retains and the publisher, by accepting the article for publication,
acknowledges that the U.S. Government retains a nonexclusive, paid-up,
irrevocable, worldwide license to publish or reproduce the published form of this
work, or allow others to do so, for U.S. Government purposes.
\end{acknowledgments}

\bibliography{references}

@article{Bardarson,
doi = {10.1088/1751-8113/41/40/405203},
url = {https://dx.doi.org/10.1088/1751-8113/41/40/405203},
year = {2008},
month = {sep},
publisher = {},
volume = {41},
number = {40},
pages = {405203},
author = {Bardarson, J H},
title = {{A proof of the Kramers degeneracy of transmission eigenvalues from antisymmetry of the scattering matrix}},
journal = {Journal of Physics A: Mathematical and Theoretical}
}

@article{Gohler2011,
author = {G{\"{o}}hler, B. and Hamelbeck, V. and Markus, T. Z. and Kettner, M. and Hanne, G. F. and Vager, Z. and Naaman, R. and Zacharias, H.},
doi = {10.1126/science.1199339},
issn = {00368075},
journal = {Science},
number = {6019},
pages = {894},
pmid = {21330541},
title = {{Spin selectivity in electron transmission through self-assembled monolayers of double-stranded DNA}},
volume = {331},
year = {2011}
}

@article{Inui2020,
author = {Inui, Akito and Aoki, Ryuya and Nishiue, Yuki and Shiota, Kohei and Kousaka, Yusuke and Shishido, Hiroaki and Hirobe, Daichi and Suda, Masayuki and Ohe, Jun-ichiro and Kishine, Jun-ichiro and Yamamoto, Hiroshi M and Togawa, Yoshihiko},
doi = {10.1103/PhysRevLett.124.166602},
issn = {1079-7114},
journal = {Phys. Rev. Lett.},
number = {16},
pages = {166602},
publisher = {American Physical Society},
title = {{Chirality-induced spin-polarized state of a chiral crystal CrNb$_3$S$_6$}},
url = {https://doi.org/10.1103/PhysRevLett.124.166602},
volume = {124},
year = {2020}
}

@article{Bloom2024,
   title = {{Chiral Induced Spin Selectivity}},
   author = "Bloom, {Brian P.} and Yossi Paltiel and Ron Naaman and Waldeck, {David H.}",
   year = "2024",
   doi = "10.1021/acs.chemrev.3c00661",
   volume = "124",
   pages = {1950},
   journal = "Chem. Rev."
}

@article{Sun2024:SciAdv,
   author = {Rui Sun and Ziqi Wang and Brian P. Bloom and Andrew H. Comstock and Cong Yang and Aeron McConnell and Caleb
                  Clever and Mary Molitoris and Daniel Lamont and Zhao-Hua Cheng and Zhe Yuan and Wei Zhang and Axel
                  Hoffmann and Jun Liu and David H. Waldeck and Dali Sun},
   title = {{Colossal anisotropic absorption of spin currents induced by chirality}},
   journal = {Sci. Adv.},
   volume = {10},
   number = {18},
   pages = {eadn3240},
   year = {2024},
   doi = {10.1126/sciadv.adn3240},
   URL = {https://www.science.org/doi/abs/10.1126/sciadv.adn3240}
}

@ARTICLE{Ray1999:Science,
   author = {K. Ray and S. P. Ananthavel and D. H. Waldeck and R. Naaman},
   year = "1999",
   journal = "Science",
   title = {{Asymmetry in the Transmission of Polarized Electrons through Organized Organic Films of Chiral Molecules}},
   volume = "283",
   pages = {814},
   doi = {10.1126/science.283.5403.8}
}

@ARTICLE{Zhao2025:NatComm,
   author = {Yufei Zhao and Kai Zhang and Jiewen Xiao and Kai Sun and Binghai Yan},
   year = "2025",
   journal = "Nat. Commun.",
   title = {{Magnetochiral charge pumping due to charge trapping and skin effect in chirality-induced spin selectivity}},
   volume = "16",
   pages = {37},
   doi = {10.1038/s41467-024-55433-1}
}

@ARTICLE{LiuYan2021,
   author = {Yizhou Liu and Jiewen Xiao and Jahyun Koo and Binghai Yan},
   year = "2021",
   journal = "Nat. Mater.",
   title = {{Chirality-driven topological electronic structure of DNA-like materials.}},
   volume = "20",
   pages = {638},
   doi = {10.1038/s41563-021-00924-5}
}

@article{Alwan2021:JACS,
author = {Alwan, Seif and Dubi, Yonatan},
title = {{Spinterface Origin for the Chirality-Induced Spin-Selectivity Effect}},
journal = {J Am. Chem. Soc.},
volume = {143},
number = {35},
pages = {14235},
year = {2021},
doi = {10.1021/jacs.1c05637},
URL = {https://doi.org/10.1021/jacs.1c05637}
}

@article{Fransson2020:PRB,
  title = {{Vibrational origin of exchange splitting and chiral-induced spin selectivity}},
  author = {Fransson, J.},
  journal = {Phys. Rev. B},
  volume = {102},
  issue = {23},
  pages = {235416},
  numpages = {8},
  year = {2020},
  month = {Dec},
  publisher = {American Physical Society},
  doi = {10.1103/PhysRevB.102.235416},
  url = {https://link.aps.org/doi/10.1103/PhysRevB.102.235416}
}

@article{Du2020:PRB,
  author  = {Du, G.-F. and Fu, H.-H. and Wu, R.},
  title   = {Vibration-enhanced spin-selective transport of electrons in the DNA double helix},
  journal = {Physical Review B},
  volume  = {102},
  pages   = {035431},
  doi     = {https://doi.org/10.1103/PhysRevB.102.035431},
  year    = {2020}
}

@article{Zhang2020:PRB,
  title = {{Chiral-induced spin selectivity: A polaron transport model}},
  author = {Zhang, Longlong and Hao, Yuying and Qin, Wei and Xie, Shijie and Qu, Fanyao},
  journal = {Phys. Rev. B},
  volume = {102},
  issue = {21},
  pages = {214303},
  numpages = {6},
  year = {2020},
  month = {Dec},
  publisher = {American Physical Society},
  doi = {10.1103/PhysRevB.102.214303},
}

@article{KleinMichaeli,
  author  = {Klein, D. and Michaeli, K.},
  title   = {Giant chirality-induced spin selectivity of polarons},
  journal = {Physical Review B},
  volume  = {107},
  pages   = {045404},
  doi     = {10.1103/PhysRevB.107.045404},
  year    = {2023}
}

@article{Sarkar2025:ACSNano,
  author = {Sarkar, Subhajit and Sharoni, Amos and Monti, Oliver L. A. and Dubi, Yonatan},
  title = {The Spinterface Mechanism for the Chiral-Induced Spin Selectivity Effect: A Critical Perspective},
  journal = {ACS Nano},
  volume = {19},
  number = {43},
  pages = {37484},
  year = {2025},
  doi = {10.1021/acsnano.5c16086}
}

@article{GuoSun,
  title = {{Spin-Selective Transport of Electrons in DNA Double Helix}},
  author = {Guo, Ai-Min and Sun, Qing-feng},
  journal = {Phys. Rev. Lett.},
  volume = {108},
  issue = {21},
  pages = {218102},
  numpages = {5},
  year = {2012},
  month = {May},
  publisher = {American Physical Society},
  doi = {10.1103/PhysRevLett.108.218102}
}

@article{GuoSunPNAS,
  author = {Ai-Min Guo and Qing-Feng Sun},
  title = {Spin-dependent electron transport in protein-like single-helical molecules},
  journal = {Proceedings of the National Academy of Sciences},
  volume = {111},
  number = {32},
  pages = {11658-11662},
  year = {2014},
  doi = {10.1073/pnas.1407716111}
}

@article{Gobel2025:arxiv,
  title = {Chirality-induced orbital Edelstein effect in an analytically solvable model},
  author = {G\"obel, B\"orge and Schimpf, Lennart and Mertig, Ingrid},
  journal = {Phys. Rev. Res.},
  volume = {7},
  issue = {3},
  pages = {033180},
  numpages = {10},
  year = {2025},
  month = {Aug},
  publisher = {American Physical Society},
  doi = {10.1103/vpjm-ntbh}
}

@article{Dong2025:AnnRevPChem,
   author = "Dong, Yifan and Hautzinger, Matthew P. and Haque, Md Azimul and Beard, Matthew C.",
   title = "Chirality-Induced Spin Selectivity in Hybrid Organic-Inorganic Perovskite Semiconductors",
   journal= "Annu.  Rev. Phys. Chem.",
   year = "2025",
   volume = "76",
   number = "Volume 76, 2025",
   pages = {519},
   doi = "https://doi.org/10.1146/annurev-physchem-082423-032933",
   url = "https://www.annualreviews.org/content/journals/10.1146/annurev-physchem-082423-032933",
   publisher = "Annual Reviews",
   issn = "1545-1593",
   type = "Journal Article",
  }

@unpublished{DiVentra2025,
  title={{Chirality-induced Spin-Orbit Coupling and Spin Selectivity}},
  author={M. Di Ventra and R. Gutierrez and G. Cuniberti},
  archivePrefix={arXiv},
  eprint={2502.09338v3}
}

@article{Xie2011,
author = {Xie, Zouti and Markus, Tal Z. and Cohen, Sidney R. and Vager, Zeev and Gutierrez, Rafael and Naaman, Ron},
title = {{Spin Specific Electron Conduction through DNA Oligomers}},
journal = {Nano Lett.},
volume = {11},
number = {11},
pages = {4652},
year = {2011},
doi = {10.1021/nl2021637}
}

@ARTICLE{Kumar2017,
author = {Anup Kumar  and Eyal Capua  and Manoj K. Kesharwani  and Jan M. L. Martin  and Einat Sitbon  and David H. Waldeck  and Ron Naaman },
title = {{Chirality-induced spin polarization places symmetry constraints on biomolecular interactions}},
journal = "Proc. Natl. Acad. Sci.",
volume = {114},
number = {10},
pages = {2474},
year = {2017},
doi = {10.1073/pnas.1611467114}
}

@article{Aiello2022:ACSNano,
author = {Aiello, Clarice D. and Abendroth, John M. and Abbas, Muneer and Afanasev, Andrei and Agarwal, Shivang and Banerjee, Amartya S. and Beratan, David N. and Belling, Jason N. and Berche, Bertrand and Botana, Antia and Caram, Justin R. and Celardo, Giuseppe Luca and Cuniberti, Gianaurelio and Garcia-Etxarri, Aitzol and Dianat, Arezoo and Diez-Perez, Ismael and Guo, Yuqi and Gutierrez, Rafael and Herrmann, Carmen and Hihath, Joshua and Kale, Suneet and Kurian, Philip and Lai, Ying-Cheng and Liu, Tianhan and Lopez, Alexander and Medina, Ernesto and Mujica, Vladimiro and Naaman, Ron and Noormandipour, Mohammadreza and Palma, Julio L. and Paltiel, Yossi and Petuskey, William and Ribeiro-Silva, João Carlos and Saenz, Juan Jos{\'e} and Santos, Elton J. G. and Solyanik-Gorgone, Maria and Sorger, Volker J. and Stemer, Dominik M. and Ugalde, Jesus M. and Valdes-Curiel, Ana and Varela, Solmar and Waldeck, David H. and Wasielewski, Michael R. and Weiss, Paul S. and Zacharias, Helmut and Wang, Qing Hua},
title = {{A Chirality-Based Quantum Leap}},
journal = {ACS Nano},
volume = {16},
number = {4},
pages = {4989},
year = {2022},
doi = {10.1021/acsnano.1c01347},
URL = {https://doi.org/10.1021/acsnano.1c01347}
}

@article{Evers2022,
author = {Evers, Ferdinand and Aharony, Amnon and Bar-Gill, Nir and Entin-Wohlman, Ora and Hedegård, Per and Hod, Oded
                  and Jelinek, Pavel and Kamieniarz, Grzegorz and Lemeshko, Mikhail and Michaeli, Karen and Mujica,
                  Vladimiro and Naaman, Ron and Paltiel, Yossi and Refaely-Abramson, Sivan and Tal, Oren and Thijssen,
                  Jos and Thoss, Michael and van Ruitenbeek, Jan M. and Venkataraman, Latha and Waldeck, David H. and
                  Yan, Binghai and Kronik, Leeor},
title = {{Theory of Chirality Induced Spin Selectivity: Progress and Challenges}},
journal = {Adv.  Mater.},
volume = {34},
number = {13},
pages = {2106629},
doi = {10.1002/adma.202106629},
url = {https://advanced.onlinelibrary.wiley.com/doi/abs/10.1002/adma.202106629},
year = {2022}
}

@unpublished{Behera2025,
  title={{Relativistic Dynamics and Electron Transport in Isolated Chiral Molecules}},
  author={Sushant Kumar Behera and Ruggero Sala and Abhirup Roy Karmakar and Matteo Moioli and Rocco Martinazzo and Matteo Cococcioni},
  archivePrefix={arXiv},
  eprint={2412.18413v2}
}

@article{Fransson2024,
author = {Fransson, J. and Turin, L.},
title = {{Current Induced Spin-Polarization in Chiral Molecules}},
journal = {J. Phys. Chem. Lett.},
volume = {15},
number = {24},
pages = {6370},
year = {2024},
doi = {10.1021/acs.jpclett.4c01362}
}

@article{Varela2012_JPCM,
doi = {10.1088/0953-8984/26/1/015008},
url = {https://doi.org/10.1088/0953-8984/26/1/015008},
year = {2013},
month = {nov},
publisher = {IOP Publishing},
volume = {26},
number = {1},
pages = {015008},
author = {Varela, Solmar and Medina, Ernesto and López, Floralba and Mujica, Vladimiro},
title = {{Inelastic electron scattering from a helical potential: transverse polarization and the structure factor in the single scattering approximation}},
journal = {Journal of Physics: Condensed Matter}
}

@article{Chiesa2024:NanoLetters,
author = {Chiesa, Alessandro and Garlatti, Elena and Mezzadri, Matteo and Celada, Leonardo and Sessoli, Roberta and Wasielewski, Michael R. and Bittl, Robert and Santini, Paolo and Carretta, Stefano},
title = {{Many-Body Models for Chirality-Induced Spin Selectivity in Electron Transfer}},
journal = {Nano Lett.},
volume = {24},
number = {39},
pages = {12133},
year = {2024},
doi = {10.1021/acs.nanolett.4c02912}
}

@article{FayLimmer,
  author  = {Fay, Thomas P. and Limmer, David T.},
  title   = {Origin of Chirality Induced Spin Selectivity in Photoinduced Electron Transfer},
  journal = {Nano Letters},
  volume  = {21},
  pages   = {6696--6702},
  doi     = {10.1021/acs.nanolett.1c02370},
  year    = {2021}
}

@article{Huisman2023:JPCC,
author = {Huisman, Karssien Hero and Heinisch, Jan-Brian Mi-Yu and Thijssen, Joseph Marie},
title = {{Chirality-Induced Spin Selectivity (CISS) Effect: Magnetocurrent–Voltage Characteristics with Coulomb Interactions I}},
journal = {J. Phys. Chem. C},
volume = {127},
number = {14},
pages = {6900},
year = {2023},
doi = {10.1021/acs.jpcc.2c08807}
}

@article{Vittmann2023,
author = {Vittmann, Clemens and Lim, James and Tamascelli, Dario and Huelga, Susana F. and Plenio, Martin B.},
title = {{Spin-Dependent Momentum Conservation of Electron–Phonon Scattering in Chirality-Induced Spin Selectivity}},
journal = {J. Phys. Chem. Lett.},
volume = {14},
number = {2},
pages = {340},
year = {2023},
doi = {10.1021/acs.jpclett.2c03224}
}

@ARTICLE{Shitade2020,
   author = {Atsuo Shitade and Emi Minamitani},
   year = "2020",
   journal = "New J. Phys.",
   title = {{Geometric spin–orbit coupling and chirality-induced spin selectivity}},
   volume = "22",
   pages = {113023},
   doi = {10.1088/1367-2630/abc920}
}

@article{Yan2024,
   author = "Yan, Binghai",
   title = "Structural Chirality and Electronic Chirality in Quantum Materials",
   journal= "Annu. Rev. Mater. Res.",
   year = "2024",
   volume = "54",
   pages = {97},
   doi = "https://doi.org/10.1146/annurev-matsci-080222-033548",
   url = "https://www.annualreviews.org/content/journals/10.1146/annurev-matsci-080222-033548",
   publisher = "Annual Reviews",
   issn = "1545-4118",
   type = "Journal Article",
  }

@article{Rikken2001,
  title = {{Electrical Magnetochiral Anisotropy}},
  author = {Rikken, G. L. J. A. and F\"olling, J. and Wyder, P.},
  journal = {Phys. Rev. Lett.},
  volume = {87},
  issue = {23},
  pages = {236602},
  numpages = {4},
  year = {2001},
  month = {Nov},
  publisher = {American Physical Society},
  doi = {10.1103/PhysRevLett.87.236602}
}

@article{Ghosh2018,
author = {Koyel Banerjee-Ghosh  and Oren Ben Dor  and Francesco Tassinari  and Eyal Capua  and Shira Yochelis  and Amir Capua  and See-Hun Yang  and Stuart S. P. Parkin  and Soumyajit Sarkar  and Leeor Kronik  and Lech Tomasz Baczewski  and Ron Naaman  and Yossi Paltiel },
title = {{Separation of enantiomers by their enantiospecific interaction with achiral magnetic substrates}},
journal = {Science},
volume = {360},
number = {6395},
pages = {1331},
year = {2018},
doi = {10.1126/science.aar4265},
URL = {https://www.science.org/doi/abs/10.1126/science.aar4265}
}

@ARTICLE{Hautzinger2024,
   author = {Matthew P. Hautzinger and Xin Pan and Steven C. Hayden and Jiselle Y. Ye and Qi Jiang and Mickey J. Wilson and Alan J. Phillips and Yifan Dong and Emily K. Raulerson and Ian A. Leahy and Chun-Sheng Jiang and Jeffrey L. Blackburn and Joseph M. Luther and Yuan Lu and Katherine Jungjohann and Z. Valy Vardeny and Joseph J. Berry and Kirstin Alberi and Matthew C. Beard},
   year = "2024",
   journal = "Nature",
   title = {{Room-temperature spin injection across a chiral perovskite/III–V interface}},
   volume = "631",
   pages = {307},
   doi = {https://doi.org/10.1038/s41586-024-07560-4}
}

@article{Eckvahl2023,
author = {Hannah J. Eckvahl  and Nikolai A. Tcyrulnikov  and Alessandro Chiesa  and Jillian M. Bradley  and Ryan M. Young  and Stefano Carretta  and Matthew D. Krzyaniak  and Michael R. Wasielewski },
title = {{Direct observation of chirality-induced spin selectivity in electron donor–acceptor molecules}},
journal = {Science},
volume = {382},
number = {6667},
pages = {197},
year = {2023},
doi = {10.1126/science.adj5328},
URL = {https://www.science.org/doi/abs/10.1126/science.adj5328}}

@ARTICLE{Sun2024,
   author = {R. Sun and K. S. Park and A. H. Comstock and A. McConnell and Y. C. Chen and P. Zhang and D. Beratan and W. You and A. Hoffmann and Z. G. Yu and Y. Diao and D. Sun},
   year = "2024",
   journal = "Nat. Mater.",
   title = {{Inverse chirality-induced spin selectivity effect in chiral assemblies of $\pi$-conjugated polymers.}},
   volume = "23",
   pages = {782},
   doi = {https://doi.org/10.1038/s41563-024-01838-8}
}

@article{Jacquod2012:PRB,
  title = {{Onsager relations in coupled electric, thermoelectric, and spin transport: The tenfold way}},
  author = {Jacquod, Philippe and Whitney, Robert S. and Meair, Jonathan and B\"uttiker, Markus},
  journal = {Phys. Rev. B},
  volume = {86},
  issue = {15},
  pages = {155118},
  numpages = {13},
  year = {2012},
  month = {Oct},
  publisher = {American Physical Society},
  doi = {10.1103/PhysRevB.86.155118}
}

@article{Kiran2016:AdvMater,
author = {Kiran, Vankayala and Mathew, Shinto P. and Cohen, Sidney R. and Hernández Delgado, Irene and Lacour, Jérôme and Naaman, Ron},
title = {{Helicenes—A New Class of Organic Spin Filter}},
journal = {Adv.  Mater.},
volume = {28},
number = {10},
pages = {1957},
doi = {https://doi.org/10.1002/adma.201504725},
year = {2016}
}

@article{Yang2019:PRB,
  title = {{Spin-dependent electron transmission model for chiral molecules in mesoscopic devices}},
  author = {Yang, Xu and van der Wal, Caspar H. and van Wees, Bart J.},
  journal = {Phys. Rev. B},
  volume = {99},
  issue = {2},
  pages = {024418},
  numpages = {16},
  year = {2019},
  month = {Jan},
  publisher = {American Physical Society},
  doi = {10.1103/PhysRevB.99.024418}
}

@article{YangVanWees,
  title = {{Reply to ``Comment on `Spin-dependent electron transmission model for chiral molecules in mesoscopic devices'''}},
  author = {Yang, Xu and van der Wal, Caspar H. and van Wees, Bart J.},
  journal = {Phys. Rev. B},
  volume = {101},
  issue = {2},
  pages = {026404},
  numpages = {3},
  year = {2020},
  month = {Jan},
  publisher = {American Physical Society},
  doi = {10.1103/PhysRevB.101.026404}
}

@article{Lu2019,
author = {Haipeng Lu  and Jingying Wang  and Chuanxiao Xiao  and Xin Pan  and Xihan Chen  and Roman Brunecky  and Joseph J. Berry  and Kai Zhu  and Matthew C. Beard  and Zeev Valy Vardeny },
title = {{Spin-dependent charge transport through 2D chiral hybrid lead-iodide perovskites}},
journal = {Sci. Adv.},
volume = {5},
number = {12},
pages = {eaay0571},
year = {2019},
doi = {10.1126/sciadv.aay0571},
URL = {https://www.science.org/doi/abs/10.1126/sciadv.aay0571},
bstract = {Chiral molecules in Pb-halide 2D-layered films achieve up to 86\% spin-polarized transport, enabling novel
                  spintronic applications. Chiral-induced spin selectivity (CISS) occurs when the chirality of the
                  transporting medium selects one of the two spin ½ states to transport through the media while blocking
                  the other. Monolayers of chiral organic molecules demonstrate CISS but are limited in their efficiency
                  and utility by the requirement of a monolayer to preserve the spin selectivity. We demonstrate CISS in
                  a system that integrates an inorganic framework with a chiral organic sublattice inducing chirality to
                  the hybrid system. Using magnetic conductive-probe atomic force microscopy, we find that oriented
                  chiral 2D-layered Pb-iodide organic/inorganic hybrid perovskite systems exhibit CISS. Electron
                  transport through the perovskite films depends on the magnetization of the probe tip and the
                  handedness of the chiral molecule. The films achieve a highest spin-polarization transport of up to
                  86\%. Magnetoresistance studies in modified spin-valve devices having only one ferromagnet electrode
                  confirm the occurrence of spin-dependent charge transport through the organic/inorganic layers.}
}

@ARTICLE{Qian2022,
   author = {Qi Qian and Huaying Ren and Jingyuan Zhou and Zhong Wan and Jingxuan Zhou and Xingxu Yan and Jin Cai and Peiqi Wang and Bailing Li and Zdenek Sofer and Bo Li and Xidong Duan and Xiaoqing Pan and Yu Huang and Xiangfeng Duan},
   year = "2022",
   journal = "Nature",
   title = {{Chiral molecular intercalation superlattices.}},
   volume = "606",
   pages = {902},
   doi = {https://doi.org/10.1038/s41586-022-04846-3}
}

@ARTICLE{Long2018,
   author = {Long and G. and Jiang and C. and Sabatini and R. and et al.},
   year = "2018",
   journal = "Nat. Photon.",
   title = {{Spin control in reduced-dimensional chiral perovskites}},
   volume = "12",
   pages = {528},
   doi = {https://doi.org/10.1038/s41566-018-0220-6}
}

@ARTICLE{KimSeebeck,
   author = {Kyunghoon Kim and Eric Vetter and Liang Yan and Cong Yang and Ziqi Wang and Rui Sun and Yu Yang and Andrew H. Comstock and Xiao Li and Jun Zhou and Lifa Zhang and Wei You and Dali Sun and Jun Liu},
   year = "2023",
   journal = "Nat. Mater.",
   title = {{Chiral-phonon-activated spin Seebeck effect.}},
   volume = "22",
   pages = {322},
   doi = {https://doi.org/10.1038/s41563-023-01473-9}
}

@article{Fransson2021:Nanolett,
author = {Fransson, Jonas},
title = {{Charge Redistribution and Spin Polarization Driven by Correlation Induced Electron Exchange in Chiral Molecules}},
journal = {Nano Lett.},
volume = {21},
number = {7},
pages = {3026},
year = {2021},
doi = {10.1021/acs.nanolett.1c00183}
}

@ARTICLE{Michaeli2017,
   author = {Karen Michaeli and Vaibhav Varade and Ron Naaman and David H Waldeck},
   year = "2017",
   journal = "J. Phys.: Condens. Matter",
   title = {{A new approach towards spintronics–spintronics with no magnets}},
   volume = "29",
   pages = {103002},
   doi = {10.1088/1361-648X/aa54a4}
}

@article{Wolf2022,
author = {Wolf, Yotam and Liu, Yizhou and Xiao, Jiewen and Park, Noejung and Yan, Binghai},
title = {{Unusual Spin Polarization in the Chirality-Induced Spin Selectivity}},
journal = {ACS Nano},
volume = {16},
number = {11},
pages = {18601},
year = {2022},
doi = {10.1021/acsnano.2c07088}
}

@article{Naaman2012:JPCL,
   author = {Ron Naaman and David H. Waldeck},
   title = {{Chiral-Induced Spin Selectivity Effect}},
   journal = {J. Phys. Chem. Lett.},
   volume = {3},
   number = {16},
   pages = {2178},
   year = {2012},
   doi = {10.1021/jz300793y}
}

@ARTICLE{Naaman2019:NatRevChem,
   author = {Ron Naaman and Yossi Paltiel and David H. Waldeck},
   year = "2019",
   journal = "Nat Rev Chem",
   title = {{Chiral molecules and the electron spin}},
   volume = "3",
   pages = {250},
   doi = {https://doi.org/10.1038/s41570-019-0087-1}
}

@article{Taguchi2018:PRB,
  title = {{Valley Edelstein effect in monolayer transition-metal dichalcogenides}},
  author = {Taguchi, K. and Zhou, B. T. and Kawaguchi, Y. and Tanaka, Y. and Law, K. T.},
  journal = {Phys. Rev. B},
  volume = {98},
  issue = {3},
  pages = {035435},
  numpages = {9},
  year = {2018},
  month = {Jul},
  publisher = {American Physical Society},
  doi = {10.1103/PhysRevB.98.035435}
}

@ARTICLE{Wang2025,
   author = {Xia Wang and Qun Yang and Sukriti Singh and Horst Borrmann and Vicky Hasse and Changjiang Yi and Yongkang Li and Marcus Schmidt and Xiaodong Li and Gerhard H. Fecher and Dong Zhou and Binghai Yan and Claudia Felser},
   year = "2025",
   journal = "Nat. Energy",
   title = {{Topological semimetals with intrinsic chirality as spin-controlling electrocatalysts for the oxygen evolution reaction}},
   volume = "10",
   pages = {101},
   doi = {https://doi.org/10.1038/s41560-024-01674-9}
}

@article{Gersten2013,
    author = {Gersten, Joel and Kaasbjerg, Kristen and Nitzan, Abraham},
    title = {{Induced spin filtering in electron transmission through chiral molecular layers adsorbed on metals with strong spin-orbit coupling}},
    journal = {J. Chem. Phys.},
    volume = {139},
    number = {11},
    pages = {114111},
    year = {2013},
    month = {09},
    issn = {0021-9606},
    doi = {10.1063/1.4820907},
    url = {https://doi.org/10.1063/1.4820907}
}

@article{Dong2025,
author = {Yifan Dong  and Aeron McConnell  and Matthew P. Hautzinger  and Md Azimul Haque  and Andrew H. Comstock  and Pius M. Theiler  and Joseph M. Luther  and Peter C. Sercel  and Dali Sun  and Matthew C. Beard },
title = {{Ultrafast inverse chirality-induced spin selectivity observed by THz emission}},
journal = {Science},
volume = {390},
number = {6773},
pages = {595},
year = {2025},
doi = {10.1126/science.ady4885},
URL = {https://www.science.org/doi/abs/10.1126/science.ady4885}
}

@ARTICLE{Cinchetti2017,
   author = {Mirko Cinchetti and V. Alek Dediu and Luis E. Hueso},
   year = "2017",
   journal = "Nat. Mater.",
   title = {{Activating the molecular spinterface}},
   volume = "16",
   pages = {507},
   doi = {https://doi.org/10.1038/nmat4902}
}

@article{Pan2025:ACSApplMat,
author = {Pan, Ruiheng and Tang, Xiantong and Wang, Xue and Liu, Yutong and Huang, Leyi and Wang, Yongjie and Wang, Zhen and Zhou, Xianju},
title = {{Impact of Chiral Spinterfaces on Magneto-Photoluminescence Effects for Chiral Lead Halide Perovskites}},
journal = {ACS Applied Materials \& Interfaces},
volume = {16},
number = {2},
pages = {2964},
year = {2024},
doi = {10.1021/acsami.3c15855}
}

@article{Xu2024:JCTC,
author = {Xu, Junqing and Ping, Yuan},
title = {{Ab Initio Predictions of Spin Relaxation, Dephasing, and Diffusion in Solids}},
journal = {J. Chem. Theory Comput.},
volume = {20},
number = {2},
pages = {492},
year = {2024},
doi = {10.1021/acs.jctc.3c00598}
}

@article{Fransson2022,
  author = {Das, Tapan Kumar and Tassinari, Francesco and Naaman, Ron and Fransson, Jonas},
  title = {{Temperature-Dependent Chiral-Induced Spin Selectivity Effect: Experiments and Theory}},
  journal = {J. Phys. Chem. C},
  volume = {126},
  number = {6},
  pages = {3257},
  year = {2022},
  doi = {10.1021/acs.jpcc.1c10550},
  URL = {https://doi.org/10.1021/acs.jpcc.1c10550}
}

@unpublished{Gupta2025:arxiv,
  title={{The role of orbital polarization and spin-dependent electron-phonon scatterings in chiral-induced spin selectivity}},
  author={Mayank Gupta and Andrew Grieder and Mayada Fadel and Jacopo Simoni and Junting Yu and Ravishankar Sundararaman and Yuan Ping},
  archivePrefix={arXiv},
  eprint={2508.03886v1}
}

@article{Fransson2023:PhysRevRes,
  title = {{Chiral phonon induced spin polarization}},
  author = {Fransson, J.},
  journal = {Phys. Rev. Res.},
  volume = {5},
  issue = {2},
  pages = {L022039},
  numpages = {6},
  year = {2023},
  month = {May},
  publisher = {American Physical Society},
  doi = {10.1103/PhysRevResearch.5.L022039}
}

@unpublished{Sato2025:arxiv,
  title={{Orbital Accumulation Induced by Chiral Phonons}},
  author={Tetsuya Sato and Takeo Kato and Aurelien Manchon},
  archivePrefix={arXiv},
  eprint={2511.11272}
}

@article{Theiler2026,
  title = {{Equilibrium spin polarization arising from chirality}},
  journal = {Newton},
  pages = {100480},
  year = {2026},
  doi = {https://doi.org/10.1016/j.newton.2026.100480},
  author = {Pius M. Theiler and Matthew C. Beard}
}

@article{Li2026,
    author = {Upadhyay, Shiv and Zheng, Xuechen and Wang, Tian and Shayit, Agam and Liu, Jun and Sun, Dali and Li, Xiaosong},
    title = {{Chirality-driven magnetization emerges from relativistic four-current dynamics}},
    journal = {APL Computational Physics},
    volume = {2},
    pages = {016103},
    year = {2026},
    doi = {10.1063/5.0313445}
}

@article{Kubo1957:JPSJ,
  author = {Kubo, Ryogo},
  title = {{Statistical-Mechanical Theory of Irreversible Processes. I. General Theory and Simple Applications to Magnetic and Conduction Problems}},
  journal = {J. Phys. Soc. Jap.},
  volume = {12},
  number = {6},
  pages = {570},
  year = {1957},
  doi = {10.1143/JPSJ.12.570},
  URL = {https://doi.org/10.1143/JPSJ.12.570}
}

@unpublished{Sala2026,
      title={{Spin-orbit coupling and beyond in Chiral-Induced Spin Selectivity}},
      author={Ruggero Sala and Sushant Kumar Behera and Abhirup Roy Karmakar and Matteo Moioli and Rocco Martinazzo and Matteo Cococcioni},
      archivePrefix={arXiv},
      eprint={2601.20475}}

@ARTICLE{Upadhyay2026,
   author = {V Upadhyay and A. Levy},
   year = "2026",
   journal = "J Chem Phys.",
   title = {{Weak electron-phonon coupling is insufficient to generate significant CISS in two-terminal transport}},
   volume = "164",
   pages = "164303",
   doi = {10.1063/5.0324706.}
}

@article{WanYan2023,
  author={L. Wan and Y.  Liu and M. J. Fuchter and B. Yan},
  title={{Anomalous circularly polarized light emission in organic light-emitting diodes caused by orbital–momentum locking}},
  journal={Nat. Photon.},
  volume={17},
  pages = {193},
  year={2023},
  doi = {10.1038/s41566-022-01113-9}
}

@article{Liu2021:NM,
  author={Y.  Liu and J. Xiao and J. Koo and B. Yan},
  title={{Chirality-driven topological electronic structure of {DNA}-like materials}},
  journal={Nat. Mater},
  volume={20},
  pages = {638},
  year={2021},
  doi = {10.1038/s41563-021-00924-5}
}

@article{Sercel2025,
  author = {Sercel, Peter C. and Hautzinger, Matthew P. and Song, Ruyi and Blum, Volker and Beard, Matthew C.},
  title = {{Optical Activity of Chiral Excitons}},
  journal = {Adv. Mater.},
  volume = {37},
  pages = {2415901},
  year = {2025},
  doi = {https://doi.org/10.1002/adma.202415901},
}

@article{Jana2020,
  author = {Jana,  Manoj K. and Song,  Ruyi and Liu,  Haoliang and Khanal,  Dipak Raj and Janke,  Svenja M. and Zhao,  Rundong and Liu,  Chi and Valy Vardeny,  Z. and Blum,  Volker and Mitzi,  David B.},
  title = {{Organic-to-inorganic structural chirality transfer in a {2D} hybrid perovskite and impact on {Rashba-Dresselhaus} spin-orbit coupling}},
  journal = {Nat. Commun.},
  volume = {11},
  pages = {4699},
  year = {2020},
  doi = {10.1038/s41467-020-18485-7},
}

@ARTICLE{Yoda2015,
   author = {T. Yoda and T. Yokoyama and S. Murakami},
   year = "2015",
   journal = "Sci. Rep.",
   title = {{Current-induced orbital and spin magnetizations in crystals with helical structure}},
   volume = "5",
   pages = {12024},
   doi = {https://doi.org/10.1038/srep12024}
}

@article{Mtangi2017,
  author = {Mtangi, Wilbert and Tassinari, Francesco and Vankayala, Kiran and Vargas Jentzsch, Andreas and Adelizzi, Beatrice and Palmans, Anja R. A. and Fontanesi, Claudio and Meijer, E. W. and Naaman, Ron},
  title = {{Control of Electrons’ Spin Eliminates Hydrogen Peroxide Formation During Water Splitting}},
  journal = {J.  Am. Chem. Soc.},
  volume = {139},
  number = {7},
  pages = {2794},
  year = {2017},
  doi = {10.1021/jacs.6b12971}
}

@article{Nair2023,
  author = {Nair, Aruna Narayanan and Fernandez, Sara and Marcos-Hernández, Mariana and Romo, Daniel Rascon and Singamaneni, Srinivasa Rao and Villagran, Dino and Sreenivasan, Sreeprasad T.},
  title = {{Spin-Selective Oxygen Evolution Reaction in Chiral Iron Oxide Nanoparticles: Synergistic Impact of Inherent Magnetic Moment and Chirality}},
  journal = {Nano Lett.},
  volume = {23},
  number = {19},
  pages = {9042},
  year = {2023},
  doi = {10.1021/acs.nanolett.3c02752}
}

@Article{Chae2024,
  author = {Chae, Kyunghee and Mohamad, Nur Aqlili Riana Che and Kim, Jeonghyeon and Won, Dong-Il and Lin, Zhiqun and Kim, Jeongwon and Kim, Dong Ha},
  title  = {{The promise of chiral electrocatalysis for efficient and sustainable energy conversion and storage: a comprehensive review of the CISS effect and future directions}},
  journal ="Chem. Soc. Rev.",
  year  = "2024",
  volume = "53",
  pages  = {9029},
  publisher = "The Royal Society of Chemistry",
  doi =  {10.1039/D3CS00316G}
}

@misc{herbrych2026:arxiv,
      title={{Electronic correlations driving Chirality-Induced Spin Selectivity}},
      author={Jacek Herbrych and Maria Daghofer},
      year={2026},
      archivePrefix={arXiv},
      eprint={2605.30240},
}

@article{Huang2026:NatPhys,
  author  = {Yijing Huang and Nick Abboud and Yinchuan Lv and Penghao Zhu and Azel Murzabekova and Changjun Lee and Emma A. Pappas and Dominic Petruzzi and Jason Y. Yan and Dipanjan Chaudhuri and Peter Abbamonte and Daniel P. Shoemaker and Rafael M. Fernandes and Jorge Noronha and Fahad Mahmood},
  title   = {{Dynamic magneto-chiral instability in photoexcited tellurium}},
  journal = {Nat. Phys.},
  volume  = {22},
  pages   = {202--208},
  year    = {2026},
  doi     = {10.1038/s41567-025-03145-8},
}

@article{Savi2025,
    author = {Savi, Lorenzo and Celada, Leonardo and Phan Huu, D.K. Andrea and Chiesa, Alessandro and Carretta, Stefano and Painelli, Anna},
    title = {{Chirality-Induced Spin Selectivity: A Minimal Model}},
    journal = {The Journal of Physical Chemistry Letters},
    volume = {16},
    number = {35},
    pages = {9107},
    year = {2025},
    month = {08},
    doi = {10.1021/acs.jpclett.5c01813}
}

@article{Slawinska2023,
    author = {Sławińska, Jagoda},
    title = {{Spin selectivity in elemental tellurium and other chiral materials}},
    journal = {Applied Physics Letters},
    volume = {123},
    number = {24},
    pages = {240504},
    year = {2023},
    month = {12},
    doi = {10.1063/5.0184391}
}

@article{MeirWingreen,
  title = {Landauer formula for the current through an interacting electron region},
  author = {Meir, Yigal and Wingreen, Ned S.},
  journal = {Phys. Rev. Lett.},
  volume = {68},
  pages = {2512},
  year = {1992},
  doi = {10.1103/PhysRevLett.68.2512}
}

@book{Datta1995,
  author    = {S. Datta},
  title     = {Electronic Transport in Mesoscopic Systems},
  publisher = {Cambridge University Press},
  year      = {1995}
}

@book{HaugJauho,
  author    = {H. Haug and A.-P. Jauho},
  title     = {Quantum Kinetics in Transport and Optics of Semiconductors},
  edition   = {2nd},
  publisher = {Springer},
  year      = {2008}
}

@book{Stefanucci2013,
  author    = {G. Stefanucci and R. van Leeuwen},
  title     = {Nonequilibrium Many-Body Theory of Quantum Systems: A Modern Introduction},
  publisher = {Cambridge University Press},
  year      = {2013}
}

@article{Lipavsky1986,
  author  = {P. Lipavsk\'y and V. \v{S}pi\v{c}ka and B. Velick\'y},
  title   = {Generalized Kadanoff-Baym ansatz for deriving quantum transport equations},
  journal = {Phys. Rev. B},
  volume  = {34},
  pages   = {6933},
  year    = {1986}
}

@article{Frederiksen2007,
  author  = {T. Frederiksen and M. Paulsson and M. Brandbyge and A.-P. Jauho},
  title   = {Inelastic transport theory from first principles: Methodology and application to nanoscale devices},
  journal = {Phys. Rev. B},
  volume  = {75},
  pages   = {205413},
  year    = {2007}
}

@article{Galperin2007,
  author  = {M. Galperin and M. A. Ratner and A. Nitzan},
  title   = {Molecular transport junctions: vibrational effects},
  journal = {J. Phys.: Condens. Matter},
  volume  = {19},
  pages   = {103201},
  year    = {2007}
}

@article{Aoki2014,
  author  = {H. Aoki and N. Tsuji and M. Eckstein and M. Kollar and T. Oka and P. Werner},
  title   = {Nonequilibrium dynamical mean-field theory and its applications},
  journal = {Rev. Mod. Phys.},
  volume  = {86},
  pages   = {779},
  year    = {2014}
}

@misc{Acharya2026Selection,
  author        = {S. Acharya and M. van Schilfgaarde},
  title         = {{Current selects the helicity of a chiral phonon}},
  year          = {2026},
  archivePrefix = {arXiv},
  eprint        = {2610.05714}
}

@article{vanSchilfgaarde2026:Newton,
  author  = {M. van Schilfgaarde and S. Acharya and K. Alberi and L. R. Baker and M. C. Beard and K. Belashchenko and J. Fransson and S. Hammes-Schiffer and M. A. Haque and M. F. Islam and A. I. Krylov and X. Li and J. M. Luther and N. T. Maitra and R. Naaman and M. R. Pederson and Y. Ping and Z. Tao and P. M. Theiler and V. Vlcek and D. Waldeck and B. Yan and I. Zutic and B. Peng and D. Sun},
  title   = {{Challenges and Opportunities in Chiral Induced Spin Selectivity}},
  journal = {Newton},
  year    = {2026},
  note    = {in press},
}

@unpublished{TheilerBeard1,
  author = {P. M. Theiler and S. Driessen and M. C. Beard},
  title  = {{Non-Hermitian Exchange as the Origin of Chirality-Induced Spin Selectivity}},
  note   = {arXiv:2505.06173},
  year   = {2025}
}

@Article{Choi2024,
  author="Choi, In Hyeok and Jeong, Seung Gyo and Song, Sehwan and Park, Sungkyun and Shin, Dong Bin and Choi, Woo Seok and Lee, Jong Seok",
  title="Real-time dynamics of angular momentum transfer from spin to acoustic chiral phonon in oxide heterostructures",
  journal="Nature Nanotechnology",
  year="2024",
  month="Sep",
  day="01",
  volume="19",
  number="9",
  pages="1277",
  doi="10.1038/s41565-024-01719-w"
}

@article{Zhu2018,
  author  = {H. Zhu and J. Yi and M.-Y. Li and J. Xiao and L. Zhang and C.-W. Yang and R. A. Kaindl and L.-J. Li and Y. Wang and X. Zhang},
  title   = {Observation of chiral phonons},
  journal = {Science},
  volume  = {359},
  pages   = {579},
  year    = {2018}
}

@article{Ueda2023,
  author  = {H. Ueda and M. Garc\'{i}a-Fern\'{a}ndez and S. Agrestini and C. P. Romao and J. van den Brink and N. A. Spaldin and Ke-Jin Zhou and U. Staub},
  title   = {Chiral phonons in quartz probed by X-rays},
  journal = {Nature},
  volume  = {618},
  pages   = {946},
  year    = {2023}
}

@article{Tauchert2022,
  author  = {S. R. Tauchert and M. Volkov and D. Ehberger and D. Kazenwadel and M. Evers and H. Lange and A. Donges and A. Book and W. Kreuzpaintner and U. Nowak and P. Baum},
  title   = {Polarized phonons carry angular momentum in ultrafast demagnetization},
  journal = {Nature},
  volume  = {602},
  pages   = {73},
  year    = {2022}
}

@Article{Trama2022,
  author = {Trama, Mattia and Cataudella, Vittorio and Perroni, Carmine Antonio and Romeo, Francesco and Citro, Roberta},
  title = {Tunable Spin and Orbital Edelstein Effect at (111) LaAlO3/SrTiO3 Interface},
  journal = {Nanomaterials},
  volume = {12},
  year = {2022},
  pages = {2494},
doi = {10.3390/nano12142494}
}

@article{Gao2022,
    author = {Gao, Mingsheng and Zhang, Kangning and Hao, Xiao-Tao and Qin, Wei},
    title = {Synergistic Effect of Chiral Nanofibers Amplifying the Orbit Angular Momentum To Enhance Optomagnetic Coupling},
    journal = {ACS Nano},
    volume = {16},
    number = {3},
    pages = {4843-4850},
    year = {2022},
    doi = {10.1021/acsnano.2c00404}
}

@article{Edelstein,
  author={V. M. Edelstein},
  title={{Spin polarization of conduction electrons induced by electric current in two-dimensional asymmetric electron system}},
  journal={Solid State Commun.},
  volume={73},
  pages={233},
  year={1990},
  doi = {https://doi.org/10.1016/0038-1098(90)90963-C}
}

@article{Johansson,
  title={{Theory of spin and orbital {Edelstein} effects}},
  author={A. Johansson},
  journal={J. Phys.: Condens. Matter},
  volume={36},
  pages={423002},
  year={2024},
  url={https://doi.org/10.1088/1361-648X/ad5e2b}
}

@Article{Roy2022,
  author="Roy, Arunesh  and Cerasoli, Frank T.  and Jayaraj, Anooja  and Tenzin, Karma  and Nardelli, Marco Buongiorno and S{\l}awi{\'{n}}ska, Jagoda",
  title={{Long-range current-induced spin accumulation in chiral crystals}},
  journal="npj Computational Materials",
  year="2022",
  month="Nov",
  day="18",
  volume="8",
  number="1",
  pages="243",
  doi="10.1038/s41524-022-00931-3",
}

@article{EndresDNA,
  title = {Colloquium: The quest for high-conductance DNA},
  author = {Endres, R. G. and Cox, D. L. and Singh, R. R. P.},
  journal = {Rev. Mod. Phys.},
  volume = {76},
  issue = {1},
  pages = {195--214},
  year = {2004},
  publisher = {American Physical Society},
  doi = {10.1103/RevModPhys.76.195}
}

@Article{WrightPhonon,
  author="Wright, Adam D.  and Verdi, Carla  and Milot, Rebecca L.  and Eperon, Giles E.  and P{\'e}rez-Osorio, Miguel A.  and Snaith, Henry J.  and Giustino, Feliciano  and Johnston, Michael B. and Herz, Laura M.",
  title={{Electron--phonon coupling in hybrid lead halide perovskites}},
  journal="Nature Communications",
  year="2016",
  volume="7",
  pages="11755",
  doi="10.1038/ncomms11755"
}

@article{Even2013,
    author = {Even, Jacky and Pedesseau, Laurent and Jancu, Jean-Marc and Katan, Claudine},
    title = {{Importance of Spin–Orbit Coupling in Hybrid Organic/Inorganic Perovskites for Photovoltaic Applications}},
    journal = {The Journal of Physical Chemistry Letters},
    volume = {4},
    pages = {2999},
    year = {2013},
    month = {08},
    doi = {10.1021/jz401532q}
}

@article{Fransson2023,
    author = {Fransson, J.},
    title = {{Vibrationally Induced Magnetism in Supramolecular Aggregates}},
    journal = {The Journal of Physical Chemistry Letters},
    volume = {14},
    number = {10},
    pages = {2558-2564},
    year = {2023},
    month = {03},
    doi = {10.1021/acs.jpclett.3c00157}
}

@misc{Kapon2026:arxiv,
      title={{Temperature-Enhanced Coercive Field by Chiral Molecules}},
      author={Yael Kapon and Lilach Brann and Shira Yochelis and Jonas Fransson and Dimitar D. Sasselov and Yossi Paltiel and S. Furkan Ozturk},
      eprint={2412.05720},
      archivePrefix={arXiv}
}

@article{Alwan2023,
    author = {Alwan, Seif and Sarkar, Subhajit and Sharoni, Amos and Dubi, Yonatan},
    title = {{Temperature-dependence of the chirality-induced spin selectivity effect—Experiments and theory}},
    journal = {The Journal of Chemical Physics},
    volume = {159},
    number = {1},
    pages = {014106},
    year = {2023},
    month = {07},
    doi = {10.1063/5.0147886}
}

@Article{Adhikari2023,
  author="Adhikari, Yuwaraj and Liu, Tianhan  and Wang, Hailong  and Hua, Zhenqi  and Liu, Haoyang  and Lochner, Eric  and Schlottmann, Pedro  and Yan, Binghai  and Zhao, Jianhua  and Xiong, Peng",
  title="Interplay of structural chirality, electron spin and topological orbital in chiral molecular spin valves",
  journal="Nature Communications",
  year="2023",
  volume="14",
  pages="5163",
  doi="10.1038/s41467-023-40884-9"
}

@article{WangMech,
  title = {Mechanical modulation of spin-polarized transport in chiral molecular junctions},
  author = {Wang, Y. and Du, W. B. and Liu, X. and Ren, J. F. and Xie, S. J. and Timm, C. and Wang, C. K. and Hu, G. C.},
  journal = {Phys. Rev. B},
  volume = {113},
  issue = {10},
  pages = {104416},
  numpages = {12},
  year = {2026},
  month = {Mar},
  publisher = {American Physical Society},
  doi = {10.1103/l56v-mspq}
}

@article{Lee2025,
  title = {{Real Space Imaging of Spin Scattering in Chirality-Induced Spin Selectivity}},
  author={Lee, Jaehyun and Lee, Sang-Hyuk and Jeong, Uiseok and JC Dalayoan, Daryll and Shin, Soobeom and Jeong, Hu Young and Jin, Hosub and Yan, Binghai and Park, Noejung and Namgung, Seon},
  journal={ACS nano},
  volume={19},
  number={50},
  pages={42046},
  year={2025},
  publisher={ACS Publications},
  url={https://pubs.acs.org/doi/10.1021/acsnano.5c10577}
}

@misc{Xu2024,
  title={Enhancement of Chirality-Induced Spin Selectivity by Strong Electron Correlations}, 
  author={Meng Xu and Yan Chen},
  year={2024},
  eprint={2405.04316},
  archivePrefix={arXiv},
  primaryClass={cond-mat.str-el},
  url={https://arxiv.org/abs/2405.04316}
}

@article{Yokoyama2025,
  author  = {T. Yokoyama},
  title   = {Phonon {E}delstein effect in chiral metals},
  journal = {Phys. Rev. B},
  volume  = {112},
  pages   = {L020406},
  year    = {2025},
  doi     = {10.1103/lsv4-z14s}
}

@article{Yokoyama2026,
  author  = {T. Yokoyama},
  title   = {Supercurrent-induced phonon angular momentum},
  journal = {Phys. Rev. B},
  volume  = {114},
  pages   = {024305},
  year    = {2026},
  doi     = {10.1103/mz1s-9z7k}
}

\end{document}